\documentclass[aps,floats]{revtex4}
\usepackage{amsmath}
\usepackage{graphicx,epsfig}

\begin{document}
\bibliographystyle {plain}

\def\oppropto{\mathop{\propto}} 
\def\opsimeq{\mathop{\simeq}}
\def\opoverderline{\mathop{\overline}}
\def\operarrow{\mathop{\longrightarrow}}
\def\opsim{\mathop{\sim}}

\def\fig#1#2{\includegraphics[height=#1]{#2}}
\def\figx#1#2{\includegraphics[width=#1]{#2}}


\title{ Finite-size scaling properties of random transverse-field Ising chains : \\
Comparison between canonical and microcanonical ensembles for the disorder } 

\author{C\'ecile Monthus}
\affiliation{Service de Physique Th\'eorique, 
Unit\'e de recherche associ\'ee au CNRS, \\
DSM/CEA Saclay, 91191 Gif-sur-Yvette, France}

\begin{abstract}

 The Random Transverse Field Ising Chain is the simplest disordered model presenting a quantum phase transition at $T=0$. We compare analytically its finite-size scaling properties in two different ensembles for the disorder (i) the canonical ensemble, where the disorder variables are independent (ii) the microcanonical ensemble, where there exists a global constraint on the disorder variables. The observables under study are the surface magnetization, the correlation of the two surface magnetizations, the gap and the end-to-end spin-spin correlation $C(L)$ for a chain of length $L$. At criticality, each observable decays typically as $e^{- w \sqrt{L}}$ in both ensembles, but the probability distributions of the rescaled variable $w$ are different in the two ensembles, in particular in their asymptotic behaviors. As a consequence, the  dependence in $L$ of averaged observables differ in the two ensembles. For instance, the correlation $C(L)$ decays algebraically as $1/L$ in the canonical ensemble, but sub-exponentially as $e^{-c L^{1/3}}$ in the microcanonical ensemble. Off criticality, probability distributions of rescaled variables are governed by the critical exponent $\nu=2$ in both ensembles, but the following observables are governed by the exponent $\tilde \nu=1$ in the microcanonical ensemble, instead of the exponent $\nu=2$ in the canonical ensemble (a) in the disordered phase : the averaged surface magnetization, the averaged correlation of the two surface magnetizations and the averaged end-to-end spin-spin correlation (b) in the ordered phase : the averaged gap. In conclusion, the measure of the rare events that dominate various averaged observables can be very sensitive to the microcanonical constraint.

\end{abstract}

\maketitle

\section{Introduction}

\subsection{ Microcanonical ensemble versus canonical ensemble for a finite disordered sample }

In the study of disordered systems, it is usual to consider
that the random variables defining
the disorder in a given sample are independent :
following \cite{igloi98,dharyoung},
this procedure will be called here the ``canonical ensemble"
 (this procedure is also called the ``grand-canonical
ensemble" in \cite{paz1,aharony,domany,paz2}). 
However, it has been argued in \cite{paz1}
that it is much more interesting in some cases
to consider the so called ``microcanonical ensemble"
as in \cite{igloi98,dharyoung} ( this
procedure is called the ``canonical
ensemble" in \cite{paz1,aharony,domany,paz2}) : 
in the microcanonical ensemble, 
there exists a global constraint on 
the random variables defining the disorder in a given sample
of $N$ sites.
The first important example concerns a pure system
 with a fraction $p \in [0,1]$ of impurities \cite{paz1}, i.e.
the disorder is characterized by a binary distribution  :
in the canonical ensemble obtained by putting 
independently on each site an impurity with probability $p$,
the total number of impurities
presents fluctuations of order $\sqrt{N}$ 
around its mean value $p N$, whereas in the microcanonical ensemble,
the total number of impurities is fixed to be exactly $p N$
and presents no fluctuations at all, i.e. the remaining
disorder only concerns the positions of the impurities.
  It is thus clear that the microcanonical ensemble 
is much less disordered that the canonical ensemble,
and indeed, it may yield
a spectacular noise reduction in numerical studies \cite{paz1,paz2}.
Another example where the ensemble dependence has been under study recently
is the Random Transverse Field Ising Chain \cite{igloi98,dharyoung},
defined by the Hamiltonian
\begin{eqnarray}
H=-\sum_i J_i \sigma_i^z \sigma_{i+1}^z - \sum_i h_i \sigma_i^x
\label{defrtfic}
\end{eqnarray}
where the couplings $J_i>0$ and the fields $h_i>0$ are random variables
(the signs can be fixed via a gauge transformation).
This model is of great interest because it is
the simplest disordered model presenting a quantum phase transition at $T=0$ :
the critical point situated at
\begin{eqnarray}
[\ln h]_{av} = [\ln J]_{av}
\label{criticality}
\end{eqnarray}
separates a ferromagnetic phase ($ [\ln h]_{av} < [\ln J]_{av} $)
from a disordered phase ($ [\ln h]_{av} > [\ln J]_{av} $).
This model can be studied in great details via a 
disorder-dependent real space RG analysis \cite{daniel},
which agrees with previous exact results \cite{mccoy,shankar}
whenever they can be compared.
In numerical studies of the quantum critical point 
on systems of linear size $L$, the usual canonical ensemble
has been compared to
the microcanonical ensemble defined by the global constraint \cite{igloi98,dharyoung}
\begin{eqnarray}
\sum_{i=1}^L ( \ln J_i - \ln h_i) =0
\label{microcriticality}
\end{eqnarray}
(with one fixed boundary condition so that there are 
exactly the same number $L$ of random bonds and random fields).
In the canonical ensemble, 
the quantity on the left in equation (\ref{microcriticality})
 presents fluctuations of order $\sqrt L$
around its mean value zero.
The samples of the microcanonical ensemble are thus `closer to
criticality' in some sense than the samples of the canonical ensemble.
So in both examples, the microcanonical constraint appears
as an opportunity to obtain numerical results
with much less fluctuations.

Now the question is : is that interesting from the
physics point of view?
On one hand, it has been strongly argued in \cite{paz1,paz2} that the microcanonical
ensemble should be preferred to the canonical ensemble,
because the latter introduces an ``extra noise" that may hid the ``intrinsic''
properties of the system.
On the other hand, it seems that the fluctuations of order
$\sqrt l$ for the sum of $l$ random variables are precisely 
an essential property of disordered systems, 
and indeed these fluctuations appear in the Harris argument \cite{harris}
on the relevance of disorder near critical points, 
in the Imry-Ma argument \cite{imryma} for random field Ising models,
and in the Chayes {\it et al.} theorem \cite{chayes}
for the bound $\nu \geq 2/d$.
Moreover, if one divides
the system of size $L$ into two halfs of size $L/2$,
each half will present fluctuations of order $\sqrt L$ 
in both ensembles :
in the canonical ensemble, these two halfs are independent,
whereas in the microcanonical ensemble,
the two halfs are completely correlated, i.e. they have 
 exactly opposite fluctuations. From this point of view, 
the microcanonical constraint can thus appear to be quite artificial
or even biased.

Of course, it seems a priori natural to expect that these two ensembles 
should be equivalent in the thermodynamic limit, 
as was shown in \cite{aharony} for the case of a random classical ferromagnet.
However, it is clear on the two examples above
that their finite-size properties can be
very different. But since the finite-size scaling theory
 of phase transitions relates the thermodynamic exponents
to finite-size effects obtained in numerical simulations,
the discussion about these two ensembles actually leads
to the general problem of the finite-size scaling theory
for disordered systems \cite{paz1,domany,paz2}.

\subsection{ Relation with the finite-size scaling theory
 for disordered systems }

To generalize the ideas of the finite-size scaling theory
developed for pure systems to the case of
disordered systems, the central question is \cite{paz1,domany,paz2} :
when the numerical simulations
provide values for some observable,
like the susceptibility $\chi(i,L)$, for various samples $(i)$
and various sizes $L$, what is the best procedure to analyze these data?
The usual procedure consists in averaging 
the observable $\chi_m(L)=<\chi(i,L)>_i$ over the samples $(i)$ at fixed size $L$ and in
analyzing the dependence in $L$ of the mean values $\chi_m(L)$
as in the finite size scaling theory of pure systems. However, as explained
in details in \cite{paz1,domany,paz2} with 
illuminating numerical examples, 
the correct procedure to analyze the data consists
in a finite-size scaling analysis for each given sample. 
In particular, one has to defined a pseudo-critical temperature
 $T_c(i,L)$ for each sample, either by the maximum of the susceptibility \cite{domany,paz2} or by a distance minimizing procedure \cite{paz2}.
Indeed, these studies have shown
that the fluctuations of $\chi(i,L)$
over the samples are mainly due to fluctuations
in $T_c(i,L)$, and that the use of the variable $(T-T_c(i,L))$ 
in the finite-size scaling analysis yields much more accurate results
(see for instance Fig.2 and Fig.3 in \cite{paz2}).
Within this point of view, the advantage of
the microcanonical ensemble with respect to the canonical ensemble
is justified if the microcanonical constraint determines $T_c(i,L)$
or at least reduces drastically its fluctuations over the samples,
as it is the case for binary distributions \cite{paz1,paz2}.
For quantum phase transitions, the pseudo-critical temperature
$T_c(i,L)$ of classical systems has to be replaced by 
the pseudo critical point \cite{paz2}.
In conclusion, the best justification for the microcanonical ensemble
 seems to be in systems where the microcanonical constraint
actually corresponds to a constraint on the pseudo-critical temperature
like in critical phenomena with a binary distribution \cite{paz1,paz2}
or to a constraint on the pseudo critical point 
for quantum phase transitions as in the Random Transverse Field Ising model.

 \subsection{ Ensemble dependence in the RTFIC  }

The ensemble dependence in the RTFIC has 
already been studied in \cite{igloi98,dharyoung}.
The first considered observable
has been the surface magnetization $m_S$
: in particular, its averaged value over the samples
was found to present different scaling in $L$ at criticality in the two ensembles \cite{igloi98,dharyoung} :
it decays as $1/{\sqrt L}$ in the canonical ensemble
and as $1/L$ in the microcanonical sample 
\cite{igloi98,dharyoung}.
However, the interpretation that should be given to
this difference is still under debate \cite{igloi98,dharyoung} :
in \cite{igloi98} it was interpreted as the presence
of the two different correlation exponents
$\nu=2$ and $\tilde \nu=1$ \cite{daniel},
whereas in \cite{dharyoung} it was interpreted
as the same exponent $\nu=2$ in both ensembles,
with an accidental vanishing of an amplitude
in the microcanonical ensemble.

In \cite{dharyoung}, other observables 
have been numerically compared at criticality
in the two ensembles, from the point of view of probability distributions
as well as averaged values.
The considered observables are
 the middle-point magnetization,
the end-to-end 
spin-spin correlation 
\begin{eqnarray}
C(L) \equiv < \sigma_1^z \sigma_L^z >
\label{defcl}
\end{eqnarray}
the gap, i.e. the energy difference between the two lowest levels
\begin{eqnarray}
\Delta(L) \equiv E_1-E_0
\label{defgapl}
\end{eqnarray}
and also 
the correlation $( m^s_1 m^s_L)$ between 
 the two surface magnetizations \cite{dharyoung}
where  
each surface magnetization  
corresponds to the magnetization at one end when the spin at the other end is fixed
\begin{eqnarray}
m^s_1 && \equiv < \sigma_1^z> \vert_{\sigma_L^z=1} \nonumber \\
m^s_L && \equiv < \sigma_L^z > \vert_{\sigma_1^z=1} 
\label{defmsms}
\end{eqnarray}
 We refer to \cite{dharyoung}
for a detailed discussion on the numerical results
for each observable.

The aim of this paper is to compare 
analytically the behavior
of these various observables in both ensembles, at criticality as well as off-criticality, with a particular attention
to the averaged values, that are governed by rare events.

The paper is organized as follows.
In Section \ref{mskesten}, 
we recall the relation between the surface magnetization
and Kesten random variables, and introduce the important notations. 
In Section \ref{saddlems}, we show that the limit
distributions for the surface magnetization
obtained in \cite{dharyoung}
can be understood as resulting from
a saddle point method in each sample.
In Section \ref{secexactmscano} and  \ref{secexactmsmicro},
we compute the exact distribution and the exact averaged value
of the surface magnetization, in the 
canonical and microcanonical ensembles respectively,
via path-integral methods. 
In Section \ref{corremsms}, we apply the saddle-point method
to compute the limit distributions
for the correlation of the two surface magnetizations
in the two ensembles.
In Section \ref{gapcorre}, we compute 
via the real-space renormalization approach
the probability distribution
of the gap and of the spin-spin end-to-end correlation 
in the microcanonical ensemble and compare with
the previously known results for the canonical ensemble \cite{rgfinitesize}.
Finally, the Section \ref{secconclusion} contains the conclusions
and the appendixes some technical details.

\section{ Surface magnetization and Kesten random variables }

\label{mskesten}

In this Section, we recall that the surface magnetization
is a Kesten random variable, and introduce
the important notations.

\subsection{ Surface magnetization in each given sample  }

Remarkably, the surface magnetization 
(\ref{defmsms}) in any given sample of size $(L+1)$
has the following exact expression \cite{igloi98,dharyoung}
\begin{eqnarray}
m_1^S= \left[ 1 + Z_L \right]^{-1/2}
\label{msdef}
\end{eqnarray}
where
\begin{eqnarray}
Z_L \equiv \sum_{i=1}^L \prod_{j=1}^i \left( \frac{h_i}{J_i} \right)^2
\label{kestenzl}
\end{eqnarray}
has a specific structure of a sum of products of random variables : 
we will call $Z_L$ a Kesten random variable \cite{kes75}.
This type of random variables also appears
in the context of random walks in random media
for various observables \cite{sol75,sinai,derrida,afa90},
as well as in the random field Ising chain
via the formulation with $2 \times 2$ random transfer matrices \cite{hilhorst,calan}.

\subsection{ Exponent $\mu$ for discrete Kesten random variables }

It is actually convenient to rewrite the Kesten random variable
(\ref{kestenzl}) with a varying left boundary
\begin{eqnarray}
Z(a,b)=\sum_{i=a}^b \prod_{j=1}^i y_j
\label{kestendiscrete}
\end{eqnarray}
where $y_j=(h_i/J_i)^2$ are the independent random variables.
The fundamental property of $Z(a,b)$ is the recurrence equation 
\begin{eqnarray}
Z(a,b)=  y_a \left[ 1+ Z(a+1,b) \right]
\label{rec}
\end{eqnarray}
where the random coefficient $ y_a$ appears multiplicatively :
$Z(a,b)$ is thus a multiplicative stochastic process.

One of the main outcome of the studies on the random variable $Z(a,b)$ 
is that, in the limit 
of infinite length $L=b-a \to \infty$, there exists a limit distribution
$P_{\infty}(Z)$ if 
\begin{eqnarray}
[\ln y ]_{av} <0
\label{lnsnegatif}
\end{eqnarray}
Moreover, the limit distribution then presents
the algebraic tail \cite{kes73,kes75,hilhorst,calan} 
\begin{eqnarray}
P_{\infty}(Z) \opsim_{Z \to \infty} \frac{1}{Z^{1+\mu} }
\label{zunplusmu}
\end{eqnarray}
where the exponent $\mu$ is defined
as the positive root $\mu>0$ of the equation
\begin{eqnarray}
[ y^{\mu} ]_{av} =1
\label{eqmudiscret}
\end{eqnarray}
In the field of random walks in random media, this exponent $\mu$
is known to govern the anomalous diffusion behavior $x \sim t^{\mu}$
in the domain $0<\mu<1$ \cite{kes75,derrida,bou90,us_sinai}.
In the context of the RTFIC, the exponent $\mu$ defined by (\ref{eqmudiscret})
has for analog  
the RG-invariant exponent $-(2 \Delta)$ \cite{igloiexpdelta}
defined by $[ (J/h)^{2 \Delta} ]_{av} = 1$, and
$z=1/(2 \Delta)$ can be interpreted as a dynamical exponent.
 In the vicinity of the critical point $\Delta=0$,
one may perform a series expansion in $ \Delta$,
to obtain the solution as $ \Delta =  \delta + O(\delta^2)$,
where
\begin{eqnarray}
 \delta = 
- \frac{ [ \ln \frac{J}{h} ]_{av} }{ [ ( \ln \frac{J}{h})^2 ]_{av}  - ([ \ln \frac{J}{h} ]_{av})^2}
=   \frac{ [ \ln h]_{av}- [ \ln J]_{av} }
{ var(\ln h) + var (\ln J) }
\label{smalldelta}
\end{eqnarray}
 is the parameter introduced in \cite{daniel}
to measure the deviation from criticality in the real-space renormalization
approach. The expression (\ref{smalldelta}), which comes from an
expansion in the two first cumulants of the variable $\ln (J/h)$,
is thus exact if the variable $\ln (J/h)$ is Gaussian, and it is only
an approximation near the critical point for all other distributions
which present higher order cumulants.

\subsection{ Continuous version of Kesten random variables}

The continuous version of the Kesten random variable (\ref{kestendiscrete})
is the exponential functional \cite{bou90,flux}
\begin{eqnarray}
Z[a,b]= \int_a^b dx e^{- \int_a^x dy F(y) }
\label{kestencontinuous}
\end{eqnarray}
where $\{F(x)\}$ is the random process 
corresponding to the random variables $(-\ln y_i)= -2 \ln (h_i/J_i)$
in the continuous limit.  The analog of the recurrence equation (\ref{rec})
is the stochastic differential equation
\begin{eqnarray}
\partial_a Z[a,b]= F(a) Z[a,b]-1 
\label{langevenmultipli}
\end{eqnarray}
where the random process $F(x)$ appears multiplicatively,
in contrast with usual Langevin equations where the noise appears additively.
In the limit 
of infinite length $L=b-a \to \infty$, 
the condition (\ref{lnsnegatif}) to have a limit distribution
$P_{\infty}(Z)$ becomes a condition on the mean
value of the process $F(x)$ that should be strictly positive 
\begin{eqnarray}
F_0 \equiv [F(x)]_{av}  >0
\label{meanpositif}
\end{eqnarray}
The exponent $\mu$ (\ref{zunplusmu})
is now determined as the root of the equation (\ref{eqmudiscret}) 
\begin{eqnarray}
[ e^{- \mu \int_0^x dy F(y) } ]_{av}  =1
\label{eqmucontinuous}
\end{eqnarray}
for arbitrary $x$ as long as the process $F(x)$
has no correlation.

It is interesting to note that the exponential functional
(\ref{kestencontinuous}) actually
determines the stationary flux $J_L$ \cite{bur92,oshanin,flux}
that exists in a given Sinai sample $[0,L]$ between two fixed concentration
$c_0$ and $c_N=0$ (i.e. particles are injected via a reservoir at $x=0$
and are removed when they arrive at the other boundary $x=N$) :
it is simply given by the inverse of
the variable $Z_L \equiv Z[0,L]$ 
\begin{eqnarray}
J_L= \frac{c_0}{Z_L}
\label{defflux}
\end{eqnarray}
In some sense, it is the simplest physical observable  
in the Sinai diffusion, as the surface magnetization is
the simplest order parameter in the RTFIC : both
can be expressed in a simple way in terms of the Kesten random variable
of the sample.

\subsection{ Case of a Brownian process}

The simplest process for $F(x)$ is of course the case where $F(x)$
is a biased Brownian motion  
\begin{eqnarray}
<F(x) >  && = F_0 \\
<F(x) F(x')>-F_0^2 && = 2 \sigma \delta(x-x')
\label{brown}
\end{eqnarray}
In this case, the exponent $\mu$ 
solution of (\ref{eqmucontinuous}) reads \cite{bou90,flux}
\begin{eqnarray}
\mu= \frac{ F_0}{  \sigma }
\label{gaussmu}
\end{eqnarray}
Here $\mu$ exactly coincides with $(-2 \delta)$ even away from criticality
, as a consequence of the Gaussian distribution.
The Brownian process actually corresponds to the fixed point of the real-space renormalization approach \cite{daniel} and can thus be used to study the universal properties near the critical point.

\section{ Limit distributions for the surface magnetization
in the large $L$ limit   }

\label{saddlems}

In this Section, we discuss the universal limit distributions
of the random variable $(\ln m_S (L))$
 in the large $L$ limit near the critical point, 
which have been obtained
in \cite{dharyoung} via an approximation of the recurrence relation
(\ref{rec}) for $\ln Z$ by an effective biased random walk 
with a reflecting boundary at the origin. 
Here we show that these analytic results \cite{dharyoung}
can actually be understood as resulting from
a saddle point method in each sample.
This point of view gives a clearer control on the 
validity of the approximation, and yields
a better understanding of the limitations
of the results obtained for the averaged
surface magnetization that is governed by rare events.
Moreover, this approach can be then generalized to obtain 
results for more complex quantities such as the correlation 
of the two surface magnetization as discussed in Section \ref{corremsms}.

\subsection{Saddle point method in each sample}

In the case of a Brownian process (\ref{brown}), the
continuous version of the Kesten process (\ref{kestencontinuous})
can be written as
\begin{eqnarray}
Z_L =  \int_0^L dx e^{- U(x)}
\end{eqnarray}
where the potential $U(x)=  \int_0^{x} dy F(y)$
is a random walk of bias $F_0$.
At the critical point $F_0=0$, this potential thus presents
fluctuations of order $\sqrt{L}$ on the interval $[0,L]$.
As a consequence, it seems natural to evaluate
this integral in the large $L$ limit by the saddle-point method  
\begin{eqnarray}
Z_L \opsimeq_{L \to \infty} e^{ E_L } Z_{valley}
\label{saddlezl}
\end{eqnarray}
where $(-E_L)<0$ is defined as the minimum reached by the process $U(x)$ 
on the interval $[0,L]$, and where $Z_{valley}$ 
represents the partition function
of an infinitely deep Brownian valley, whose probability
distribution was studied in \cite{us_golosov}.
At the critical point, the scaling $E_L \sim \sqrt{L}$
shows that there will exists a limit distribution
for $(\ln Z_L)/\sqrt L \sim \ln m^s_L/\sqrt L$,
and that this limiting distribution is given by
the limit distribution of $E_L/\sqrt{L}$.
In the following, we study the distribution of the minimum $E_L$
at criticality and off criticality.

To study the probability distribution in the two ensembles,
we thus need the following standard result :
the probability for the biased random walk to go from $U_0 $ to $U$
during $L$ 
in the presence of an absorbing boundary at $U=U_a$ reads from
the methods of images 
\begin{eqnarray}
G^{(F_0)}_{[U_a,+\infty[}(U,L \vert U_0) = \frac{1}{\sqrt{4 \pi \sigma L} }
\left( e^{- \frac{(U-U_0-F_0 l)^2}{4  \sigma L} }
- e^{-  \frac{F_0 }{ \sigma} (U_0-U_a) }
 e^{- \frac{(U+U_0-2U_a-F_0 L)^2}{4  \sigma L} }  \right)
\label{imagesuu0}
\end{eqnarray}
The joint probability of the end-point $U$
and of the minimum value $U_{min}=-E<0$ when starting at $U_0=0$
then reads
\begin{eqnarray}
P_L^{(F_0)}(U,E) && = 
- \left[\partial_{ U_a} 
G^{(F_0)}_{[U_a,+\infty[}(U,L \vert 0)  
 \right]_{U_a=-E} \nonumber \\
&& = \theta(E>0) \theta(U>-E) \frac{ (U+2 E) }{2 \sqrt{\pi} (\sigma L)^{3/2} } 
 e^{-  \frac{F_0 }{ \sigma} E } e^{- \frac{(U+2E-F_0 L)^2}{4  \sigma L} } 
\label{loijointeue}
\end{eqnarray}

\subsection{Distribution of $E_L$ in the canonical ensemble}

In the canonical ensemble, we need the partial law of $E$
when integrating over the end-point $U$ in (\ref{loijointeue})
\begin{eqnarray}
\left[ P_L^{(F_0)}( E ) \right]_{cano} && = 
\int dU P_L^{(F_0)}(U,E)
\nonumber \\
&& =\theta(E>0)  e^{-  \frac{F_0 }{ \sigma} E } \int_{E}^{+\infty} dV
 \frac{ V }{2 \sqrt{\pi} (\sigma L)^{3/2} } 
 e^{- \frac{(V-F_0 L)^2}{4  \sigma L} }
\end{eqnarray}
The final result reads 
after an integration by part and with the notation $F_0=\sigma \mu$
(\ref{gaussmu})
\begin{eqnarray}
{\cal P}_L^{cano}(E) = \theta(E>0)
\left[ \frac{ 1 }{ \sqrt {\pi  \sigma L} } 
e^{- \frac{(E+\mu \sigma L)^2}{4  \sigma L} }  
+ \mu e^{- \mu  E }
\int_{\frac{(E+ \mu \sigma L)}{ \sqrt{4  \sigma L}}}^{+\infty} \frac{dz}{\sqrt{\pi}} e^{-z^2} \right]
\label{rescanoe}
\end{eqnarray}
which exactly coincides with the formula (10) in \cite{dharyoung}
with the correspondence  $\mu=-2 \overline{\delta}$
and $l=\sigma L$.
Let us first consider 
the limit $L \to \infty$ with $\mu$ fixed :

For $\mu>0$, there exists a limit distribution for $E$
which is simply exponential
\begin{eqnarray}
{\cal P}^{(\mu>0)}_{cano}(E) \opsimeq_{L \to \infty}  \theta(E>0)
 \mu e^{- \mu  E }
\label{canofplus}
\end{eqnarray}
This distribution actually
 corresponds to the distribution of barriers against the bias
\cite{feigelman,us_sinai} in a Sinai potential.

For $\mu \leq 0$, 
the appropriate rescaled variable reads
\begin{eqnarray}
{\cal E} \equiv \frac{(E- \vert \mu \vert \sigma L)}{ \sqrt{4  \sigma L}}
\label{rescaledE}
\end{eqnarray}
For $\mu < 0$, the rescaled variable ${\cal E}$
is distributed with the full Gaussian distribution
\begin{eqnarray}
P^{(\mu<0)}_{cano}({\cal E})=  \frac{1}{\sqrt \pi} e^{-{\cal E}^2}
\end{eqnarray}
whereas for $\mu=0$, the rescaled variable
is distributed with the half Gaussian distribution  
\begin{eqnarray}
P^{(F_0=0)}_{cano}({\cal E})= \theta({\cal E}>0) 
\frac{2}{\sqrt \pi} e^{-{\cal E}^2}
\label{canofzero}
\end{eqnarray}

\subsection{Distribution of $E_L$ in the microcanonical ensemble}

In the presence of the microcanonical constraint (\ref{micro}),
 which fixes the end-point value of the potential $U(L)=F_0L$, 
we need the distribution of $E$ 
for the trajectories having exactly $U=F_0 L$,
i.e. we obtain from the joint distribution (\ref{loijointeue})
with the notation $F_0=\mu \sigma$
\begin{eqnarray}
\left[ P_L^{(\mu)}( E ) \right]_{micro} && = 
\frac{ P_L^{(F_0=\mu \sigma)}(U=0,E) } 
{ \int dE P_L^{(F_0=\mu \sigma)}(U=0,E) } 
=
\theta(E>0) \theta(E>-\mu \sigma L ) 
\left( \mu + \frac{ 2 E }{ \sigma L } \right)
 e^{-  \mu E } e^{- \frac{E^2}{  \sigma L} }
\label{resmicroe}
\end{eqnarray}
which exactly coincides with the formula (11) in \cite{dharyoung}
with the correspondence  $\mu=-2 \overline{\delta}$
and $l=\sigma L$.
We now consider
the limit $L \to \infty$ with $\mu$ fixed, to compare with
the results of the canonical ensemble.

For $\mu >0$, there exists a limit distribution
that exactly coincides with (\ref{canofplus})
\begin{eqnarray}
{\cal P}_{micro}^{(\mu>0)}(E)  \opsimeq_{L \to \infty}
\theta(E >0)  \mu  e^{- \mu  E } 
\label{microfplus} 
 \end{eqnarray}

At the critical point $\mu = 0$, the appropriate rescaled variable
is again ${\cal E}=E/\sqrt{4  \sigma L}$,
but the corresponding distribution
\begin{eqnarray}
P^{(\mu=0)}_{micro}({\cal E}) && = \theta( {\cal E} >0) 
  8 {\cal E}   
e^{- 4 {\cal E}^2 } 
\label{microfzero}
\end{eqnarray}
does not coincide with the canonical result (\ref{canofzero}).
In particular, it vanishes linearly as ${\cal E} \to 0$,
in contrast with the canonical result which presents a finite
density at the origin. 

For $\mu<0$, the appropriate rescaled variable
is not ${\cal E}$ (\ref{rescaledE}) in contrast with the canonical case,
but it is
\begin{eqnarray}
 v \equiv E- \vert \mu \vert \sigma L 
\label{variablev}
\end{eqnarray}
which is asymptotically distributed
with the exponential distribution
\begin{eqnarray}
P^{(\mu<0)}_{micro}(v) && = \theta( v >0) \mu  
e^{- \mu  E }
\label{microfmoins}
\end{eqnarray}
which is exactly the distribution of barriers against the bias
(\ref{canofplus},\ref{microfplus}) : this shows that the 
statistics for the barrier
$E=U(0)-U(x_0)$ is actually completely determined
 in this case by the statistics
of the barrier $v=U(L)-U(x_0)$ distributed with (\ref{microfmoins}),
since they are completely correlated via the microcanonical constraint
$U(L)-U(0)=v-E=F_0 L$ (\ref{micro}).

\subsection{ Discussion   }

We first have to discuss the validity of the saddle-point method.
At the critical point $\mu = 0$, where the random variable $E_L$
 scales as $\sqrt L$, and in the phase $\mu<0$, 
where the random variable $E_L$ is of order $L$,
the saddle-point method is thus well justified at large $L$,
at least for typical samples.
On the contrary, in the phase $\mu>0$, where the random variable $E_L$
remains finite, the saddle-point method will be
a good approximation only if $E$ is large, i.e. in the limit $\mu \to 0$
near the critical point.  

On the other hand, since whenever the saddle-point analysis is valid,
the variable $E_L$ is large, we may safely approximate
the surface magnetization by 
\begin{eqnarray}
\ln m^s_1 = \ln (1+Z_L)^{-1/2} \opsimeq_{Z_L >>1 } -(1/2) \ln Z_L \sim -E_L/2
\label{drop1}
\end{eqnarray}
and thus the results derived above for the distribution of $E_L$
indeed represents the probability distribution of the log
of the surface magnetization.
In particular, in the vicinity of the critical point,
if one introduces the scaling variables
\cite{daniel,dharyoung}
\begin{eqnarray}
w && = \frac{E}{ \sqrt{  \sigma L}} 
= \frac{ -2 \ln m^s_1 }{ \sqrt{  \sigma L}}   \\
\gamma && = \mu {\sqrt {  \sigma L }}
\label{defwgamma}
\end{eqnarray}
we may write the following finite-size scaling forms
for the probability distribution of $(-\ln m^s_1)$
\begin{eqnarray}
P_L( -\ln m^s_1) = \frac{2}{ \sqrt{  \sigma L}} Q 
\left( w  = \frac{ -2 \ln m^s_1 }{ \sqrt{  \sigma L}}
 ; \gamma  = \mu {\sqrt { \sigma L }} \right)
\end{eqnarray}
where the scaling functions respectively read
for the two ensembles (\ref{rescanoe},\ref{resmicroe})
\begin{eqnarray}
Q_{cano} 
( w  ; \gamma  )
&& =  \theta(w>0)
\left[   \frac{ 1 }{ \sqrt {\pi  } } 
e^{- \frac{(w+\gamma)^2}{4 } }  
+ \gamma e^{- \gamma w } 
\int_{\frac{w-\gamma}{2}}^{+\infty} \frac{dz}{\sqrt{\pi}} e^{-z^2} \right] \\
Q_{micro} 
\left( w  ; \gamma  \right)
&& = \theta(w >0) \theta(w> - \gamma )
  \left( 2 w+\gamma  \right) 
e^{- \gamma w - w^2 } 
\label{reswithrescaling}
\end{eqnarray}
As emphasized in \cite{dharyoung}, 
the critical exponents are the same in the two ensembles
 (the scaling variables $(w,\gamma)$ are the same),
but the finite-size scaling functions are different in the two ensembles.

We now turn to the question of the mean surface magnetization.
As stressed in \cite{igloi98,dharyoung}, since the typical 
surface magnetization decays in the critical region as 
\begin{eqnarray}
m^s_1 = e^{ - \frac{\sqrt {  \sigma L }}{2} w }
\end{eqnarray}
where $w$ is a random variable of order one, 
the mean surface magnetization will be governed by 
the rare samples that presents an anomalously big
surface magnetization of order one, and its decay with $L$
will be governed by the measure of these rare samples as a function of $L$.
However, since the limit distributions above are a priori not expected
to describe well the tail $w \sim 1/\sqrt{L}$ outside the scaling region,
and since the approximations (\ref{drop1}) are not valid anymore
for $Z_L \sim 1$, it is necessary to discuss separately the samples having
$Z_L \sim 1$ from the starting point (\ref{saddlezl}) 
of the saddle-point analysis.   
For these samples, the minimum $(-E_L)$ of the potential $U(x)$
on the interval $[0,L]$ has to remain of order one.
At the critical point $\mu=0$, 
the decay of the mean surface magnetization in the canonical ensemble
\cite{igloi98,dharyoung} 
\begin{eqnarray}
[m^s_1]_{cano} \oppropto_{L \to \infty} \frac{1}{\sqrt L}
\end{eqnarray}
corresponds to the scaling of the probability 
for a Brownian path to remain 
above its starting point during a length $L$.
In the microcanonical ensemble, the presence of the different decay
\cite{igloi98,dharyoung} 
\begin{eqnarray}
[m^s_1]_{micro} \oppropto_{L \to \infty} \frac{1}{ L}
\end{eqnarray}
can be understood as the ratio between 
(i) the scaling as $1/L^{3/2}$ for the probability 
of the first return to the origin after a length $L$,
(ii) the scaling as $1/\sqrt{L}$ for the probability
to be at the origin at $L$.
These two scalings, reflecting the measure of the rare samples
having $Z_L \sim 1$, can actually be recovered from
the scaling functions (\ref{reswithrescaling})
as computed in \cite{dharyoung}, but
the numerical prefactors given in equations (12-13) of \cite{dharyoung}
should not be considered as exact. 
Actually these prefactors are not expected to be universal,
as was discussed for other quantities such as the average end-to-end correlation
in \cite{rgfinitesize}.  

In conclusion of this Section, 
we have shown that 
the limit distributions for the log of the surface magnetization
obtained in \cite{dharyoung}
can be derived via a saddle point method in each sample,
and
have thus a very simple probabilistic interpretation :
they describe the probability distribution
of the minimum of a finite-size biased random walk,
with different boundary conditions
in the canonical and in the microcanonical ensembles.
These limit distributions for the log of Kesten random variables
should be considered as the analog
of the Central-limit theorem for the log of a product of random variables
\cite{redner} : they describe well typical samples
and are expected to be universal. 
However, the computation of some averaged quantities 
which are dominated by rare events
cannot be computed exactly from
these limit distributions in the scaling regime. 
In the two next Section, we thus compute exactly
the average surface magnetization in the two ensembles.

\section{ Exact results for the surface magnetization in the canonical ensemble}

  \label{secexactmscano}

\subsection{ Exact expression for the averaged surface magnetization }

The probability distribution of 
the random variable $Z_L$ (\ref{kestencontinuous})
in the case where the process $\{F(x)\}$ is a Brownian motion (\ref{brown})
has been already much studied 
by various methods \cite{flux,yor,yorbook}.
Here, the most convenient starting point is the 
following exact result for its Laplace transform 
derived in \cite{flux}
\begin{eqnarray}
[ e^{-p Z_L } ]_{cano} = 
&&  \sum_{0 \leq n <\frac{\mu}{2} }
 e^{-\sigma L n (\mu-n) } 
\frac{ 2 (\mu-2n) }
{\Gamma(1+n) \Gamma(1+\mu-n)} \left( \frac{p}{\sigma} \right)^{\frac{\mu}{2}}
K_{\mu-2n} \left( 2 \sqrt{\frac{p}{\sigma}} \right) 
\nonumber \\
&& +\frac{ 1} { 2 \pi^2 } 
\int_0^{+\infty} dq e^{-\frac{\sigma L}{4} (\mu^2+q^2) }
 q \sinh \pi q 
\left \vert \Gamma \left( -\frac{\mu}{2} +i \frac{q}{2} \right) 
\right \vert^2
\left( \frac{p}{\sigma} \right)^{\frac{\mu}{2}}
K_{iq} \left( 2 \sqrt{\frac{p}{\sigma}} \right)
\label{genecano}
\end{eqnarray}
where $\mu$ is the exponent (\ref{gaussmu}), and where $\sigma$
represents the strength of the disorder (\ref{brown}).
We refer the interested reader to the previous studies \cite{flux,yor}
for a detailed discussion of this result.
Here, we are interested into the surface   
magnetization $m_1^s$ which can be computed from 
the result (\ref{genecano}) via the identity
\begin{eqnarray}
[m^s_1 ]_{cano} = [(1+Z_L)^{-1/2} ]_{cano} = \frac{1}{\sqrt \pi }
 \int_0^{+\infty} dp p^{-1/2} e^{-p} [ e^{-p Z_L} ]_{cano}
\end{eqnarray}
which yields in terms of the
the Whittaker function (\ref{whit})
\begin{eqnarray}
&& [m^s_1 ]_{cano} = 
  \frac{ \sigma^{\frac{1}{2}} }{ \sqrt \pi}
\sum_{0 \leq n <\frac{\mu}{2} }
 e^{-\sigma L n (\mu-n) } 
(\mu-2n) \frac{ \Gamma \left( n+ \frac{1}{2} \right)
\Gamma \left( \mu+\frac{ 1}{2}-n \right)  }
{\Gamma(1+n) \Gamma(1+\mu-n)}
\left[ \sigma^{-\frac{\mu}{2}}  e^{\frac{1 }{ 2 \sigma }}
W_{-\frac{\mu}{2} ,  \frac{\mu}{2}-n } 
\left(  \frac{1 }{  \sigma } \right) \right]
\nonumber \\
&& +\frac{  \sigma^{\frac{1}{2}} }  
 { 4 \pi^{5/2} } 
\int_0^{+\infty} dq e^{-\frac{\sigma L}{4} (\mu^2+q^2) }
 q \sinh \pi q 
\left \vert \Gamma \left( -\frac{\mu}{2} +i \frac{q}{2} \right) 
\right \vert^2
\left \vert
\Gamma \left( \frac{\mu}{2} +i \frac{q}{2}
+\frac{ 1}{2} \right) \right \vert^2
\left[ \sigma^{-\frac{\mu}{2}} e^{\frac{1 }{ 2 \sigma }} W_{-\frac{\mu}{2} , i \frac{q}{2} } 
\left(  \frac{1 }{  \sigma } \right) \right]
\label{canosurf}
\end{eqnarray}

As a comparison, the negative moment of order $(1/2)$ reads
\begin{eqnarray}
[ Z_L^{- \frac{1}{2} } ]_{cano} 
&& = \frac{ \sigma^{\frac{1}{2}} } { \sqrt \pi }
 \sum_{0 \leq n <\frac{\mu}{2} }
 e^{-\sigma L n (\mu-n) } (\mu-2n)
\frac{ \Gamma(\frac{1}{2}+n) \Gamma(\frac{1}{2}+\mu-n) }
{\Gamma(1+n) \Gamma(1+\mu-n)} 
\nonumber \\
&& +\frac{ \sigma^{\frac{1}{2}} } { 4 \pi^{5/2} } \int_0^{+\infty} dq e^{-\frac{\sigma L}{4} (\mu^2+q^2) }
 q \sinh \pi q 
\left \vert \Gamma \left( -\frac{\mu}{2} +i \frac{q}{2} \right) 
\right \vert^2
\left \vert \Gamma \left( \frac{1}{2}+\frac{\mu}{2} +i \frac{q}{2} \right) \right \vert^2
\label{moment12cano}
\end{eqnarray}
i.e. the two results are very similar : the only differences
are in the the factors inside $[..]$ in (\ref{canosurf}).

\subsection{ Ordered phase $\mu=-2 \delta>0$  }

For $\mu>0$, there exists a limit distribution in the large $L$ limit,
whose generating function is simply given by the term $n=0$ in
(\ref{genecano}) 
\begin{eqnarray}
[ e^{-p Z_{\infty} } ]_{cano} = 
\frac{ 2  }
{ \Gamma(\mu)} \left( \frac{p}{\sigma} \right)^{\frac{\mu}{2}}
K_{\mu} \left( 2 \sqrt{\frac{p}{\sigma}} \right)
\label{canolimitmuplus} 
\end{eqnarray}
It corresponds to the probability distribution \cite{bou90,flux}
\begin{eqnarray}
{\cal P} ( Z_{\infty} ) = 
\frac{\sigma}{\Gamma(\mu) } \left( \frac{1}{\sigma Z_{\infty}} \right)^{1+\mu} e^{- \frac{1}{\sigma Z_{\infty}} } 
\end{eqnarray}
For the RTFIC, the case $\mu>0$
indeed corresponds to the ordered phase $\delta<0$ where
the surface magnetization remains finite in the thermodynamic limit
\begin{eqnarray}
[ m_s(\infty) ]_{cano} (\mu>0) = 
&& \mu   \sqrt{\sigma }  
\frac{  \left(\Gamma(\frac{1}{2}+\mu) \right)  }
{ \Gamma(1+\mu)}
\left[ \sigma^{\frac{-\mu}{2}}  e^{\frac{1 }{ 2 \sigma }}
W_{-\frac{\mu}{2} ,  \frac{\mu}{2} } 
\left(  \frac{1 }{  \sigma } \right)  \right]
\label{canosurfmuplus}
\end{eqnarray}
As a comparison, we have  
\begin{eqnarray}
[ Z_{\infty}^{-\frac{1}{2}} ]_{cano} = 
&& \mu \sqrt{\sigma } 
\frac{  \Gamma(\frac{1}{2}+\mu) }
{ \Gamma(1+\mu)}
\end{eqnarray}

\subsection{ Critical point $\mu=0$ }

At the critical point $\mu=0$, the result (\ref{genecano})
for the Laplace transform
of the probability distribution
of $Z_L$ simplifies into
\begin{eqnarray}
[ e^{-p Z_L } ]_{cano} = 
\frac{ 2} {  \pi } 
\int_0^{+\infty} dq e^{-\frac{\sigma L}{4} q^2 }
  \cosh \pi \frac{q}{2} 
K_{iq} \left( 2 \sqrt{\frac{p}{\sigma}} \right)
\label{genecanocriti}
\end{eqnarray}
and the result (\ref{canosurf}) for the average surface magnetization
becomes 
using (\ref{whitzero}) 
\begin{eqnarray}
[m^s_1 ]_{cano} (\mu=0) && = 
\frac{    e^{\frac{1 }{ 2 \sigma }}}
 {  \pi } 
\int_0^{+\infty} dq e^{-\frac{\sigma L}{4} q^2 }
    K_{i \frac{q}{2}} \left( \frac{1}{2 \sigma} \right) \\
&& =
\frac{    e^{\frac{1 }{ 2 \sigma }}}
 {  \sqrt{ \pi \sigma L}  } K_0 \left( \frac{1}{2 \sigma} \right)
\left[ 1 - \frac{ K_0'' \left( \frac{1}{2 \sigma} \right) }{ 4 \sigma L 
K_0 \left( \frac{1}{2 \sigma} \right)} +O(1/L^3) \right]
\label{canosurfmuzero}
\end{eqnarray}
whereas 
the negative moment of order $(1/2)$ reads
\begin{eqnarray}
[ Z_L^{- \frac{1}{2} } ]_{cano} (\mu=0)
= \frac{ 1 } {  \sqrt{L} } 
\label{moment12canozero}
\end{eqnarray}
Both results are indeed dominated by rare configurations
of measure $1/\sqrt{ L}$ which have a $Z_L$ of order $1$,
in agreement with the previous studies \cite{igloi98,dharyoung}.
The different prefactors in the leading terms
of (\ref{canosurfmuzero},\ref{moment12canozero}) 
can be understood via the series expansion 
\begin{eqnarray}
(1+Z_L)^{-1/2} = Z_L^{-1/2} - (1/2) Z_L^{-3/2} + ...
\end{eqnarray}
Since the averages are dominated
by the rare events where $Z_L$ is of order 1 with probability of order $1/\sqrt{L}$,
all the terms in the series expansion
will contribute to the prefactor found in (\ref{canosurfmuzero}).

It is actually interesting to consider the distribution of $Z_L$ 
among these rare events : the Laplace transform
has for leading term as $L \to \infty$ with $p$ fixed
\begin{eqnarray}
[ e^{-p Z_L } ]_{cano} \opsimeq_{L \to \infty} 
\frac{ 2} { \sqrt{ \pi \sigma L}  }  
K_{0} \left( 2 \sqrt{\frac{p}{\sigma}} \right)  
+ O \left( \frac{1}{L^{3/2} } \right)
\label{genecanocritirarel}
\end{eqnarray}
which corresponds after Laplace inversion to the following 
behavior for the probability distribution $P_L(z)$
at large $L$ with fixed $z$
\begin{eqnarray}
 P_L^{cano}(z)  \opsimeq_{L \to \infty} 
\frac{ 1 } { \sqrt{ \pi \sigma L}  }  
  \frac{ e^{- \frac{1}{\sigma z } } }{ z }  
+ O \left( \frac{1}{L^{3/2} } \right)
\label{probacanocritirare}
\end{eqnarray}
This result describes the tail of rare events with $z$ fixed
outside the scaling region $\ln z \sim \sqrt{L}$ studied in 
Section \ref{saddlems}.

\subsection{ Finite-size scaling function in the critical region  }

In the critical region parameterized by the rescaled parameter $\gamma$ (\ref{defwgamma}),
we obtain that the distribution of $Z_L$ of order one 
is characterized by the Laplace transform
\begin{eqnarray}
[ e^{-p Z_L } ]_{cano} (\gamma) =   
\frac{ 2  }
{ { \sqrt { \pi \sigma L} }  } 
K_{0} \left( 2 \sqrt{\frac{p}{\sigma}} \right)
\left[ \theta( \gamma>0) \gamma {\sqrt \pi}
+  e^{- \frac{\gamma^2}{4} } -  \vert \gamma \vert 
\int_{\frac{ \vert \gamma \vert}{2} }^{+\infty} dv e^{-v^2}   \right]
+ O\left(\frac{1}{L} \right)
\end{eqnarray}
where the term containing the theta function comes from
the bound state $n=0$, and the other terms from the continuum.
Actually, we may rewrite for arbitrary sign of $\gamma$
\begin{eqnarray}
[ e^{-p Z_L } ]_{cano} (\gamma) =   
\frac{ 2  }
{ { \sqrt { \pi \sigma L} }  } 
K_{0} \left( 2 \sqrt{\frac{p}{\sigma}} \right)
\left[  e^{- \frac{\gamma^2}{4} } +  \gamma  
\int_{-\frac{  \gamma }{2} }^{+\infty} dv e^{-v^2}   \right]
+ O\left(\frac{1}{L} \right)
\label{genecanogamma}
\end{eqnarray}
This result is thus completely factorized
into (i) the distribution of $z$ among the rare events
at the critical point (\ref{genecanocritirarel})
times (ii) the function of $\gamma$ inside $[..]$
that coincides with eq (12) of \cite{dharyoung}
and that represents the probability of $E_L=0$ in the scaling function
discussed in previous Section \ref{saddlems}.
As a consequence, the finite-size scaling 
form for the averaged magnetization reads
\begin{eqnarray}
[m^s_1 ]_{cano} (\gamma) \opsimeq_{L \to \infty}    
\frac{ c_m  }
{ { \sqrt { \pi \sigma L} }  } 
\left[  e^{- \frac{\gamma^2}{4} } +  \gamma  
\int_{-\frac{  \gamma }{2} }^{+\infty} dv e^{-v^2}   \right]
+ O\left(\frac{1}{L} \right)
\label{meanmscanogamma}
\end{eqnarray}
where $c_m$ is a non-universal constant, found here to be $c_m= e^{\frac{1 }{ 2 \sigma }}
  K_0 \left( \frac{1}{2 \sigma} \right)$ (\ref{canosurfmuzero}),
and where all dependence in $\gamma$ corresponds to the prediction
of the scaling regime as derived in \cite{dharyoung}.
We now compute the same observables for
the microcanonical ensemble.

\section{ Exact results for the surface magnetization 
in the microcanonical ensemble}

  \label{secexactmsmicro}

In this Section, we derive exact results for
 the continuous Kesten variable with a Gaussian disorder
in the presence of the microcanonical constraint 
\begin{eqnarray}
 \int_0^L dy F(y)  = F_0 L = \mu \sigma L
\label{micro}
\end{eqnarray}
and we compare with the results of the canonical ensemble
given in the previous Section.

\subsection{ Probability distribution  }

The Laplace transform of the probability distribution of $Z_L$
in the microcanonical ensemble
may be written as follows in terms of path-integrals
\begin{eqnarray}
[ e^{-p Z_L}  ]_{micro}
= \frac{ \int {\cal D} F(x) e^{- \frac{1}{4 \sigma} \int_0^L [F(x)-F_0]^2
-p \int_0^L dx e^{-  \int_0^x dy F(y) }  }
\delta \left( \int_0^L dx F(x) -F_0 L \right)  }
{ \int {\cal D} F(x) e^{- \frac{1}{4 \sigma} \int_0^L [F(x)-F_0]^2  }
\delta \left( \int_0^L dx F(x) -F_0 L \right)  }
\end{eqnarray}
where the delta function represents the microcanonical constraint
(\ref{micro}).
Rewriting this path-integral over the force $F(x)$ as a path-integral
over the potential $U(x)=\int_0^x dy F(y)$ as in \cite{flux}, we obtain
\begin{eqnarray}
[ e^{-p Z_L}  ]_{micro}
= \frac{
\int_{U(0)=0}^{U(L)=F_0 L} {\cal D} U(x) e^{- \frac{1}{4 \sigma} \int_0^L \left( \frac{dU(x)}{dx} \right)^2
-p \int_0^L dx e^{ - U(x) } } }
{\int_{U(0)=0}^{U(L)=F_0 L} {\cal D} U(x) e^{- \frac{1}{4 \sigma} \int_0^L \left( \frac{dU(x)}{dx} \right)^2 } }
\end{eqnarray}
The denominator is a simple Gaussian propagator
\begin{eqnarray}
\int_{U(0)=0}^{U(L)=F_0 L} {\cal D} U(x) e^{- \frac{1}{4 \sigma} \int_0^L \left( \frac{dU(x)}{dx} \right)^2 } = \frac{1}{ \sqrt{4 \pi  \sigma L} }
e^{- \frac{ (U(L)-U(0))^2}{4 \sigma L} } = 
\frac{1}{ \sqrt{4 \pi  \sigma L} } e^{- \frac{F_0^2}{4 \sigma} L}
\end{eqnarray}

The path-integral in the numerator 
is equivalent to a propagator of quantum mechanics
corresponding to the Hamiltonian $H=-\sigma d^2/dU^2+p e^{-U}$.
Expanding this path-integral into the eigenstates $\psi_k(U)$
of the Hamiltonian as described in \cite{flux},
we finally get with $F_0=\mu \sigma$
\begin{eqnarray}
[ e^{-p Z_L}  ]_{micro}
&& =  \sqrt{ 4 \pi  \sigma L} e^{  \frac{ \mu^2}{ 4} \sigma L}
\int_{-\infty}^{+\infty} \frac{dk}{2 \pi} e^{- k^2 L} 
\psi_k^*(0) \psi_k ( \mu \sigma L) \nonumber \\
&& 
= \frac{ 2  }{ \pi^{3/2} } \sqrt{   \sigma L}
e^{ \frac{ \mu^2}{ 4} \sigma L }
\int_{0}^{+\infty} dq  e^{- \frac{q^2}{4} \sigma L}
 q \sinh \pi q
K_{i q } \left( 2 \sqrt{\frac{p}{\sigma} } \right)
K_{ i q } \left( 2 \sqrt{\frac{p}{\sigma} } 
e^{- \frac{ \mu \sigma L }{2}  } \right)
\label{genemicro}
\end{eqnarray}
which should be compared with (\ref{genecano}).

\subsection{ Surface magnetization  }

The surface   
magnetization $m_1^s$ which can then be computed  via the identity
\begin{eqnarray}
[m^s_1 ]_{micro} = [(1+Z_L)^{-1/2} ]_{micro} = \frac{1}{\sqrt \pi }
 \int_0^{+\infty} dp p^{-1/2} e^{-p} [ e^{-p Z_L} ]_{micro}
\end{eqnarray}
which yields 
\begin{eqnarray}
[m^s_1 ]_{micro}  = 
\frac{ 2  }{ \pi^2 } \sqrt{   \sigma L}
e^{ \frac{ \mu^2}{ 4} \sigma L }
\int_{0}^{+\infty} dq  e^{- \frac{q^2}{4} \sigma L}
 q \sinh \pi q
 \int_0^{+\infty} dp p^{-1/2} e^{-p}
K_{i q } \left( 2 \sqrt{\frac{p}{\sigma} } \right)
K_{ i q } \left( 2 \sqrt{\frac{p}{\sigma} } 
e^{- \frac{ \mu \sigma L }{2}  } \right)
\label{microosurf}
\end{eqnarray}

\subsection{ Ordered phase $\mu=-2 \delta>0$  }

For $\mu>0$, we have seen in the canonical ensemble that there
exists a limit distribution in the large $L$ limit,
given by (\ref{canolimitmuplus}).
To recover this result in the microcanonical ensemble,
we may use (\ref{kiqzto0}) at lowest order
to obtain
\begin{eqnarray}
K_{ i q } \left( 2 \sqrt{\frac{p}{\sigma} } 
e^{- \frac{ \mu \sigma L }{2}  } \right) && = \frac{\pi}{2 i \sinh \pi q}
 \left[\frac{\left( \frac{p}{\sigma} \right)^{-i\frac{q}{2}}
}{\Gamma(1-iq)} e^{iq \frac{ \mu \sigma L }{2}  }
-\frac{\left( \frac{p}{\sigma}\right)^{i\frac{q}{2}}
}{\Gamma(1+iq)} e^{- iq\frac{ \mu \sigma L }{2}  }\right]+
O\left(  e^{-  \mu \sigma L   } \right)
\label{kiqlowestorder}
 \end{eqnarray}
which yields
\begin{eqnarray}
[ e^{-p Z_\infty}  ]_{micro}
&& = \lim_{L \to \infty} 
\left( \frac{ \sqrt{   \sigma L}  }{ 2 i \pi^{1/2} } 
\int_{-\infty}^{+\infty} dq   q 
K_{i q } \left( 2 \sqrt{\frac{p}{\sigma} } \right)
 \left[\frac{\left( \frac{p}{\sigma} \right)^{-i\frac{q}{2}}
}{\Gamma(1-iq)} 
e^{- \frac{(q-i \mu)^2}{4} \sigma L}
-\frac{\left( \frac{p}{\sigma}\right)^{i\frac{q}{2}}
}{\Gamma(1+iq)} 
e^{- \frac{(q+i \mu)^2}{4} \sigma L}\right] \right) \nonumber \\
&& = \frac{ 2  }
{ \Gamma(\mu)} \left( \frac{p}{\sigma} \right)^{\frac{\mu}{2}}
K_{\mu} \left( 2 \sqrt{\frac{p}{\sigma}} \right)
\end{eqnarray}
where the final limit is obtained by taking into account the
saddles in the complex plane at $q=\pm i \mu$ respectively for the two terms.
The limit distribution in the thermodynamic limit is thus
the same as in the canonical ensemble (\ref{canolimitmuplus}).

\subsection{ Critical point  $\mu=0$ }

At the critical point $\mu=0$, the probability distribution
of $Z_L$ is characterized by the Laplace transform
\begin{eqnarray}
[ e^{-p Z_L}  ]_{micro} (\mu=0)
&& 
= \frac{2}{\pi^{3/2} } \sqrt{   \sigma L }
\int_{0}^{+\infty} dq  e^{- \frac{\sigma L}{4} q^2 }
 q \sinh \pi q
K_{i q }^2 \left( 2 \sqrt{\frac{p}{\sigma} } \right)
\end{eqnarray}
which should be compared with (\ref{genecanocriti}).
The surface magnetization (\ref{microosurf}) reads
\begin{eqnarray}
[m^s_1 ]_{micro}  = 
&& \frac{ 2  }{ \pi^2 } \sqrt{   \sigma L}
\int_{0}^{+\infty} dq  e^{- \frac{q^2}{4} \sigma L}
 q \sinh \pi q
 \int_0^{+\infty} dp p^{-1/2} e^{-p}
K_{i q }^2 \left( 2 \sqrt{\frac{p}{\sigma} } \right) \\
&& = \frac{ 8 \pi }{ \sigma L} 
\int_0^{+\infty} dp p^{-1/2} e^{-p}
K_{0 }^2 \left( 2 \sqrt{\frac{p}{\sigma} } \right) + 
O \left( \frac{1}{L^2} \right)
\label{microosurfmu0}
\end{eqnarray}
As a comparison, the negative moment of order $(1/2)$ reads
\begin{eqnarray}
[ Z_L^{-1/2}  ]_{micro} (\mu=0)
=  \pi  \sigma  \sqrt{ L}  
\int_{0}^{+\infty} dq  e^{- q^2 \frac{\sigma L}{4}}
q \tanh(\pi q)  =   \frac{2 \pi^{5/2}}{ L \sqrt{\sigma} }
\left[ 1- \frac{2 \pi^2}{\sigma L} +O(1/L^2)  \right]  
\end{eqnarray}
Both are indeed dominated by the rare events where $Z_L \sim 1$,
which have a measure of order $1/L$ in the microcanonical ensemble,
in agreement with previous studies \cite{igloi98,dharyoung}.

It is now interesting to consider the distribution of $Z_L$ 
for these rare events : the Laplace transform
has for leading term as $L \to \infty$ with $p$ fixed
\begin{eqnarray}
[ e^{-p Z_L } ]_{micro} (\mu=0)
\opsimeq_{L \to \infty} 
\frac{ 4} {  \sigma L  }  
K_{0}^2 \left( 2 \sqrt{\frac{p}{\sigma}} \right)  
+ O \left( \frac{1}{L^{2} } \right)
\label{genemicrocritirarel}
\end{eqnarray}
This leading behavior is actually very simply
related to the analog result (\ref{genecanocritirarel})
in the canonical ensemble
\begin{eqnarray}
[ e^{-p Z_L } ]_{micro}^{rare} \simeq 
\pi \left( [ e^{-p Z_L } ]_{cano}^{rare}  \right)^2
\label{raremicrocano}
\end{eqnarray}
which corresponds after Laplace inversion to 
a convolution for the probability distribution $P_L^{cano}(z)$
at large $L$ with fixed $z$ (\ref{probacanocritirare})
\begin{eqnarray}
 P_L^{micro}(z)  \opsimeq_{L \to \infty}
 \frac{2 }{ \sigma L z } K_0 \left( \frac{2}{\sigma z} \right) e^{- \frac{2}{\sigma z} }
\label{probamicrocritirare}
\end{eqnarray}
This result describes the tail of rare events with $z$ fixed
in the microcanonical ensemble,
outside the scaling region $\ln z \sim \sqrt{L}$ studied in 
Section \ref{saddlems}.

\subsection{ Finite size scaling function in the critical region }

In the critical region inside the ordered phase $\mu=-2 \delta >0$,
parameterized by the rescaled parameter $\gamma$ (\ref{defwgamma}),
the analog of (\ref{kiqlowestorder}) reads with $q=k/\sqrt{\sigma L}$
\begin{eqnarray}
K_{ i q } \left( 2 \sqrt{\frac{p}{\sigma} } 
e^{- \frac{ \gamma \sqrt{\sigma L} }{2}  } \right) && = \frac{\pi}{2 i \sinh \pi q}
 \left[\frac{\left( \frac{p}{\sigma} \right)^{-i\frac{q}{2}}
}{\Gamma(1-iq)} e^{iq \frac{ \gamma \sqrt{\sigma L} }{2}  }
-\frac{\left( \frac{p}{\sigma}\right)^{i\frac{q}{2}}
}{\Gamma(1+iq)} e^{- iq\frac{ \gamma \sqrt{\sigma L} }{2}  }\right]+
O\left(  e^{-  \gamma \sqrt{\sigma L}   } \right)
\label{kiqordered}
 \end{eqnarray}
which yields for the generating function with fixed $Z$ at large $L$
 \begin{eqnarray}
[ e^{-p Z_L}  ]_{micro} (\gamma=\mu \sqrt{
\sigma L}>0)
= \gamma \frac{ 2  }
{ \sqrt{ \sigma L} } 
K_{0} \left( 2 \sqrt{\frac{p}{\sigma}} \right)
+ O \left( \frac{1}{L} \right)
\end{eqnarray}
The finite-size scaling for the average surface magnetization is thus simply
\begin{eqnarray}
[m^s_1 ]_{micro} (\gamma>0) \opsimeq_{L \to \infty} 
 \frac{ \gamma c_m }{ \sqrt{ \sigma L} }  
\label{meanmsmicrogamma}
\end{eqnarray}
that should be compared with the corresponding result in the canonical ensemble 
(\ref{meanmscanogamma}) : 
$c_m$ is the same constant, and the scaling function of $\gamma$
has been simply replaced by the factor $\gamma$ alone, 
that again corresponds to the prediction
of the scaling regime as derived in \cite{dharyoung}.
Here, the amplitude of the leading term in $1/\sqrt{L}$ vanishes 
at the critical point $\gamma=0$, as it should
to recover the scaling $1/L$ at the critical point.

In the critical region inside the disordered phase $\mu=-2 \delta <0$,
parameterized by the rescaled parameter $\gamma$ (\ref{defwgamma}),
the argument in the Bessel function 
is now exponentially large in $\sqrt{L}$ instead of exponentially small 
as in (\ref{kiqordered}). Taking into account 
the asymptotic behavior of Bessel function at large argument (\ref{knulargez}) 
we thus obtain that $\gamma$ is {\it not} the appropriate scaling variable
near the critical point on the disordered side.
Instead, to obtain the finite-size scaling function,
we need the scaling variable
\begin{eqnarray}
\rho \equiv \mu \sigma L
\end{eqnarray}
that leads to
\begin{eqnarray}
[ e^{-p Z_L}  ]_{micro} (\rho = \mu \sigma L <0 )
\opsimeq_{ L \to \infty}
 \frac{ 4} {  \sigma L  }  
K_{0} \left( 2 \sqrt{\frac{p}{\sigma}} \right)
K_{0} \left( 2 \sqrt{\frac{p}{\sigma}} e^{- \frac{\rho}{2} } \right)  
+ O \left( \frac{1}{L^2 } \right)
\label{genemicrocritidisorderrare}
\end{eqnarray}
After Laplace inversion, the probability distribution $P_L(z)$
at large $L$ with fixed $z$ and $\rho$ is given by
\begin{eqnarray}
 P_L^{micro}(z,\rho)  \opsimeq_{L \to \infty} 
\frac{ 1 } {  \sigma L z }  
 \int_0^1 \frac{du}{u (1-u) }  
e^{- \frac{1}{\sigma z } 
\left( \frac{1}{u} + \frac{e^{-\rho} }{1-u} \right) }   
+ O \left( \frac{1}{L^2 } \right)
\label{probamicrocritidisorderrare}
\end{eqnarray}
So here, in contrast with the corresponding result (\ref{genecanogamma})
in the canonical ensemble, this result is not factorized 
into a function of $z$ times a function of $\rho$.
The finite-size scaling form for the average surface magnetization is thus given by
\begin{eqnarray}
[m^s_1 ]_{micro} (\rho= \mu \sigma L<0) \opsimeq_{L \to \infty}    
 \frac{ 4} { {\sqrt \pi} \sigma L  } \int_0^{+\infty} dp p^{-1/2} e^{-p} 
K_{0} \left( 2 \sqrt{\frac{p}{\sigma}} \right)
K_{0} \left( 2 \sqrt{\frac{p}{\sigma}} e^{- \frac{\rho}{2} } \right)  
+ O \left( \frac{1}{L^2 } \right)
\label{meanmsmicrorho}
\end{eqnarray}

In conclusion, the critical regime on the disordered side
 is thus governed by the scaling variable
$ \rho= \mu \sigma L <0 $ in the microcanonical ensemble, 
whereas it is governed by the scaling variable
$ \gamma=\mu \sqrt{\sigma L}<0$ in the canonical ensemble.
To understand these results, it is thus useful to recall the origin of the presence
of two different correlation length exponents in the RTFIC.

\subsection{ Discussion on the two correlation length exponents}
 
The presence of two different correlation length exponents $\nu=2$
and $\tilde{\nu}=1$ in the RTFIC has been discussed in detail in
\cite{daniel}.
The definition of these two length scale for the Brownian process $(U(L)-U(0))$ 
can be summarized as follows \cite{daniel} :
the first length scale corresponds to the length ${ \tilde \xi}$
where the mean value $<U(L)-U(0)>=F_0 L=\mu \sigma L$ is of order one, which yields
\begin{eqnarray}
{ \tilde \xi } \sim \frac{ 1}{\sigma \mu^{\tilde \nu} } \ \ \ {\rm with} \ \ {\tilde \nu}=1
\label{deftildexi}
\end{eqnarray}
The second length scale 
corresponds to the length $ \xi$
where most of the samples indeed have 
$(U(L)-U(0))$ of the same sign of the mean value,
i.e. the scale $\sqrt{ \sigma L}$ of the fluctuations 
should be of the same order of the mean value, which yields
\begin{eqnarray}
 \xi \sim \frac{ 1}{\sigma \mu^{\nu}} \ \ \ {\rm with} \ \  \nu=2
\label{defxi}
\end{eqnarray}
These two length scales appear in various quantities in the RTFIC.
In particular, the averaged correlation
is governed by the exponent $\nu=2$,
whereas the typical correlation is governed by
the exponent ${\tilde \nu}=1$ \cite{daniel}.
More generally, the presence of these two length-scales 
is also well known in related models,
for instance in  
the eigenstates of the Fokker-Planck operator in a Sinai potential
\cite{bou90,us_golosov}
and in one-dimensional random-hopping Hamiltonian for fermions
\cite{balents_fisher,eigenhuse}.

In view of this knowledge, the interpretation of our results
obtained in this Section for the averaged surface magnetization is as follows :
in the canonical ensemble, the critical behavior is governed by the
exponent $\nu=2$, whereas in the microcanonical ensemble,
there are two different exponents in the two sides of the critical point :
the ordered phase is governed by the exponent $\nu=2$,
with an accidental vanishing of the amplitude at the critical point,
as stressed in \cite{dharyoung}, but the disordered phase is governed
by the exponent ${\tilde \nu}=1$, in agreement with the analysis given in
\cite{igloi98} and in contrast with
the interpretation given in \cite{dharyoung}.
More generally, the appearance of
these two exponents for some given observable in the two ensembles
should not be too surprising : indeed,  
the requirement that almost all samples of size $L$ 
 indeed ``know'' the sign of $F_0$ involves the
length-scale $\xi$ (\ref{defxi}) in the canonical ensemble,
whereas it involves the length $ \tilde \xi$ (\ref{deftildexi})
in the microcanonical ensemble. So the very definitions
of the two length scales $\xi$ and $ \tilde \xi$
show that the microcanonical constraint can indeed play
an important role for the critical exponents. 

\subsection{ Obtaining the exponent $\tilde \nu$ from the scaling regime   }

As a final remark, it is now useful to discuss how the
exponent ${\tilde \nu}=1$, found in this Section by an exact path-integral method
for the averaged surface magnetization, can be understood from
the knowledge of the scaling regime (\ref{reswithrescaling}),
that a priori only contains the exponent $\nu=2$. 
The evaluation of the averaged surface magnetization
from the scaling function (\ref{reswithrescaling}) 
in the disordered phase 
\begin{eqnarray}
[m^s_1 ]_{micro}^{scaling} (\mu<0) = \int_0^{+\infty} dw Q_{micro} 
\left( w  ; \gamma  \right) e^{- \frac{\sqrt{\sigma L}}{2} w }
\label{nutildefromnu}
\end{eqnarray}
shows that the origin of the ``anomalous" exponent
${\tilde \nu}=1$ can be traced back
to the presence of the theta function $\theta(w> - \gamma )$
in the scaling function $Q_{micro} \left( w  ; \gamma  \right)$,
i.e. the probability density $Q_{micro} \left( w  ; \gamma  \right)$
vanishes at a finite positive value in $w$ in the disordered phase.
This remark will be useful in the next Sections on 
more complicated observables, where we will compute 
the probability distributions in the scaling regime,
and where direct methods to compute exactly the averaged values 
such as the path-integral method used in this Section will not be available.

\section{ Correlation between the two surface magnetizations
in the two ensembles }

\label{corremsms}

In this Section, we consider
the correlation $(m^s_1 m^s_L)$ (\ref{defmsms}) between 
 the two surface magnetizations 
that has been numerically studied in \cite{dharyoung}.
We compute its universal limit distributions at large $L$ in the two 
ensembles.

\subsection{ Saddle-point method in each sample}

The correlation $(m^s_1 m^s_L)$ can be expressed in closed form
in an arbitrary sample as the surface magnetization (\ref{msdef})
\begin{eqnarray}
m^s_1 m^s_L= \left[ 1 + Z_L \right]^{-1/2} \left[ 1 + {\tilde Z}_L \right]^{-1/2}
\label{mscorredef}
\end{eqnarray}
where $Z_L$ and ${\tilde Z}_L$ are two Kesten variables
defined in terms of the same random potential $U(x)=\int_0^x F(y)dy$
 in the continuous version (\ref{kestencontinuous})
\begin{eqnarray}
Z_L && = \int_0^L dx e^{U(0)- U(x) } \\
{\tilde Z}_L && = \int_0^L dx e^{ U(x)-U(L) } 
\label{twokesten}
\end{eqnarray}
 
The saddle-point method in a given sample (\ref{saddlezl})
on the associated random potential 
 $U(x)=  \int_0^{x} dy F(y)$ leads to  
\begin{eqnarray}
Z_L && \opsimeq_{L \to \infty} e^{ U(0)-U_{min} } Z_{valley}  \\
{\tilde Z}_L && \opsimeq_{L \to \infty} e^{ U_{max}-U(L) } {\tilde Z}_{valley}
\label{zzsaddle}
\end{eqnarray}
 in terms of the minimum $U_{min}$ and maximum $U_{max}$
 reached by the process $U(x)$ 
on the interval $[0,L]$.
Since both Kesten variables are typically large in the critical
region, we may replace as in (\ref{drop1})
\begin{eqnarray}
\ln ( m^s_1 m^s_L ) \opsimeq_{L \to \infty} - \frac{1}{2} 
\ln Z_L - \frac{1}{2} {\tilde Z}_L 
\opsimeq_{L \to \infty} - \frac{ U(0)-U_{min} + U_{max}-U(L)  }{2} 
\label{lnmsmswithd}
\end{eqnarray}
so the limit distribution of $\ln (m^s_1 m^s_L)$
can be computed in terms of 
the distribution of the random variable
\begin{eqnarray}
D_L \equiv  A(L)-U(L)
\label{defdl}
\end{eqnarray}
where $A(L)$ is the amplitude of the Brownian trajectory for $0 \leq x \leq L$
\begin{eqnarray}
A(L) \equiv U_{max}-U_{min}
\end{eqnarray}
and where $U(0)=0$ is the starting point and $U(L)$ the final point.

\subsection{ Joint distribution of the amplitude $A$ and the end-point $U$ }

The joint probability $P_L^{(\mu)}(A,U)$ of the amplitude $A_L=U_{max}-U_{min}=U_b-U_a$
and the end-point $U$ when starting at $U_0=0$
has for Laplace transform with respect to $L$ (see Appendix \ref{appjointau})
\begin{eqnarray}
&& {\hat P}^{(\mu)}(A,U;p)   \equiv \int_0^{+\infty} dL e^{-p L} P_L^{(\mu)}(A,U)
\nonumber \\
&& =  \theta(A>\vert U \vert) 
  e^{ \frac{\mu}{2} U }
\frac{ \cosh qU \left( q (A- \vert U \vert) \coth qA -1 \right) + \sinh q \vert U \vert \coth qA }
{ \sigma \sinh^2 q A} 
\label{loijointeulamplilaplace}
\end{eqnarray}
where
\begin{eqnarray}
q \equiv \sqrt{\frac{p}{\sigma}+\frac{\mu^2}{4} }
\end{eqnarray}
The Laplace inversion yields
\begin{eqnarray}
 P_L^{(\mu)}(A,U) && = \theta(A>\vert U \vert) 
 \frac{e^{- \frac{ (U- \mu \sigma L)^2}{4 \sigma L}} }
{2 \sqrt{ \pi } (\sigma L)^{3/2} }
\sum_{k=-\infty}^{+\infty} e^{- k^2 \frac{A^2}{  \sigma L} }
  [ (A- \vert U \vert )
 k^2 \left( \frac{(U-2 k A )^2}{\sigma L} -2 \right)
e^{ k \frac{ A U }{ \sigma L} }
\nonumber \\
&&
- 2 k (k-1)  
( \vert U \vert - 2 k A) 
e^{ k \frac{ A \vert U \vert}{ \sigma L} }   ]
\label{loijointeulampli}
\end{eqnarray}

We now discuss the microcanonical ensemble and the
canonical ensemble respectively.

\subsection{ Microcanonical ensemble }

In the microcanonical ensemble with the constraint $U(L)=\mu \sigma L$, 
the probability distribution
of the amplitude $A_L=U_{max}-U_{min}=U_b-U_a$
reads  (\ref{loijointeulampli})
\begin{eqnarray}
\left[ P_L^{(\mu)}(A) \right]_{micro} && = 
\frac{ P_L^{(\mu)}(A,\mu \sigma L) }{ \int dA  P_L^{(\mu)}(A,\mu \sigma L)}
=  
 \theta(A> \vert \mu \vert \sigma L)
  \frac{ 1 }
{  (\sigma L) }
\sum_{k=-\infty}^{+\infty} e^{- k^2 \frac{A^2}{  \sigma L} } \nonumber  \\
&&   \left[ (A- \vert \mu \vert \sigma L )
 k^2 \left( \frac{(\mu \sigma L-2 k A )^2}{\sigma L} -2 \right)
e^{ k A \mu  }
- 2 k (k-1)  
( \vert \mu \vert \sigma L - 2 k A) 
e^{ k A  \vert \mu \vert }   \right]
\end{eqnarray}
Since $D=A-\mu \sigma L$,
we may now write the final results 
for the various cases $\mu=0$, $\mu>0$ and $\mu<0$.

\subsubsection{ Critical case $\mu=0$ }

In the critical case, the probability distribution of
the variable $D=A$ reads
\begin{eqnarray}
\left[ P_L^{(\mu=0)}(D) \right]_{micro} 
= \theta( D>0) \frac{ 4 D} {\sigma L} 
\sum_{k=1}^{+\infty} k^2 \left( 2 k^2 \frac{  D^2} {\sigma L} -3 \right)
 e^{- k^2 \frac{  D^2} {\sigma L}}
\end{eqnarray}

We thus obtain that the distribution
of the variable $\ln (m_1^s m_L^s)$ (\ref{lnmsmswithd})
takes the scaling form
\begin{eqnarray}
\left[ {\cal P}_L( - \ln (m_1^s m_L^s) ) \right]_{micro} 
= \frac{2} { \sqrt{ \sigma L} } {\cal P}_{micro}^{(0)} \left( {\cal D} = \frac{-2 \ln (m_1^s m_L^s)} { \sqrt{ \sigma L} } \right)
\end{eqnarray}
with the scaling function
\begin{eqnarray}
 P_{micro}^{(0)}( {\cal D} ) 
&&  = \theta({\cal D}> 0) 
  4 {\cal D} 
\sum_{k=1}^{+\infty} k^2 ( 2 k^2 {\cal D}^2 -3 )
 e^{- k^2  {\cal D}^2 }   \\
&& = \theta({\cal D}> 0) \frac{4 \sqrt \pi }{ {\cal D}^4 }
\sum_{n=1}^{+\infty} n^2 \pi^2 \left(   \frac{ 2 n^2 \pi^2 }{{\cal D}^2 } -3\right) 
e^{- n^2 \frac{ \pi^2 }{{\cal D}^2 } } 
\end{eqnarray}
These two series representations 
are useful to study
the asymptotic behaviors at small and large arguments respectively. 
The decay at large ${\cal D}$ is given by 
\begin{eqnarray}
P_{micro}( {\cal D} )  
\opsimeq_{D \to \infty}  8  {\cal D}^3 
 e^{- {\cal D}^2 }
\end{eqnarray}
and at the origin, there is an essential singularity
\begin{eqnarray}
&& P_{micro}( {\cal D} ) 
\opsimeq_{D \to 0} \frac{8  \pi^{9/2} }{ {\cal D}^6 }
e^{-  \frac{ \pi^2 }{{\cal D}^2 } } 
\label{microcald}
\end{eqnarray}
This behavior of the scaling function at small argument
yields the following dependence in $L$ for the averaged
 correlation of the two surface magnetizations
\begin{eqnarray}
\left[ m^s_1 m^s_L \right]_{micro}^{scaling}
 \equiv \int_0^{+\infty} d {\cal D} P_{micro}( {\cal D} ) e^{ -
 {\cal D}  \frac{  \sqrt{ \sigma L} } {2}  } \oppropto_{L \to \infty} L^{2/3} 
e^{- \frac{3}{2} \left( \frac{\pi^2}{2} \sigma L \right)^{1/3} } 
\label{averagedmsmsmicro} 
\end{eqnarray}
that should be compared
with the numerical fits discussed in \cite{dharyoung}.

\subsubsection{ Ordered phase $\mu>0$}

In the ordered phase $\mu>0$, there exists
a limit probability distribution
for $D$ as $L \to \infty$ 
\begin{eqnarray}
\left[ P_{\infty}^{(\mu>0)}(D) \right]_{micro}  = 
    \theta( D  > 0 )  
   \mu^2 D 
 e^{-   \mu D }
\label{respdmicromuplus}
\end{eqnarray}
This result corresponds to the convolution of two independent barriers 
against the bias as it should.

\subsubsection{ Disordered phase $\mu<0$}

For $\mu<0$, the minimal value of $D$ is $D_{min}=- 2 \mu  \sigma L$,
so it is convenient to set
\begin{eqnarray}
D \equiv  2 \vert \mu \vert \sigma L +W
\end{eqnarray}
and there exists a limit distribution for $W$
in the large $L$ limit that reads
\begin{eqnarray}
\left[ P_{L \to \infty}^{(\mu<0)}(W) \right]_{micro}  =   
 \theta(W > 0) \mu^2 W e^{- \vert W \vert }
 \end{eqnarray}
This results simply represents the distribution
of the sum of two barriers against the bias :
the comparison with the result (\ref{microfmoins})
concerning one surface magnetization
shows that it can be interpreted in the same way.

\subsubsection{ Finite-size scaling function}

Finally, in the vicinity of the critical point,
we may write the following finite-size scaling form
for the probability distribution of $(-\ln m^s_1 m^s_L)$
\begin{eqnarray}
P_L( -\ln m^s_1 m^s_L) = \frac{2}{ \sqrt{  \sigma L}} {\cal Q} 
\left( {\cal D}  = \frac{ -2 \ln m^s_1  m^s_L }{ \sqrt{  \sigma L}}
 ; \gamma  = \mu {\sqrt { \sigma L }} \right)
\end{eqnarray}
where the scaling function reads
\begin{eqnarray}
&& {\cal Q} \left( {\cal D}  
 ; \gamma  \right)
= 
 \theta( {\cal D} > 0)
   \theta( {\cal D} >  - 2 \gamma)
 \sum_{k=-\infty}^{+\infty} e^{- k^2 ( {\cal D} + \gamma)^2 } \nonumber   \\
&&   \left[ k^2 ({\cal D} + \gamma- \vert \gamma \vert  )
  \left( 2 k {\cal D} + (2 k-1) \gamma \right)^2
e^{ k \gamma ( {\cal D} + \gamma)  }
+ 2 k (k-1)  
( 2 k ( {\cal D} + \gamma) -\vert \mu \vert \sigma L ) 
e^{ k \vert \gamma \vert ( {\cal D} + \gamma) }   \right]
\label{scalingmsms}
\end{eqnarray}

By comparison with the results obtained before for the surface magnetization,
we expect that the averaged correlation of the two surface magnetizations
is governed by the exponent $\nu=2$ in the ordered phase
and by the exponent ${\tilde \nu}=1$ in the disordered phase, 
for the reasons explained around Equation (\ref{nutildefromnu}) :
the scaling function (\ref{scalingmsms}) is defined in terms of
the exponent $\nu=2$, but the presence of the theta function $\theta( {\cal D} >  - 2 \gamma)$
that forbids a finite interval in ${\cal D}$ near the origin
in the disordered phase
yields that the averaged value $[m^s_1 m_s^L ]_{micro}$ will be 
governed by the exponent ${\tilde \nu}=1$.

\subsection{ Canonical ensemble }

For the canonical ensemble,
it is more convenient to characterize 
the probability distribution
of $D$ by its Laplace with respect to $L$ using
the result (\ref{loijointeulamplilaplace})

\begin{eqnarray}
&& \left[ {\hat P}^{(\mu)}(D;p) \right]_{cano}=\int_0^{+\infty} dL e^{-p L}
\left[ P_L^{(\mu)}(D) \right]_{cano}
= \int dA \int dU {\hat P}^{(\mu)}(A,U;p) \delta( D-( A-U) )
\nonumber \\
&& = \theta(D>0) D \int_1^{+\infty} dv  
  e^{ \frac{\mu}{2} D (v-1) }
\frac{ q D \cosh q D v \cosh q D (v-1)    - \sinh q D }
{ \sigma \sinh^3 q D v} \\
&& + \theta(D>0) D \int_{1/2}^1 dv
   e^{ \frac{\mu}{2} D (v-1) }
\frac{  q D (2 v-1) \cosh q D v \cosh q D (1-v) 
+  \sinh q D (1-2 v)   }
{  \sigma \sinh^3 q Dv}
\end{eqnarray}

In particular at criticality $\mu=0$,
 the distribution
of the variable $D$ 
takes the scaling form
\begin{eqnarray}
\left[ {\cal P}_L( D) ) \right]_{cano} 
= \frac{1} { \sqrt{ \sigma L} } {\cal P}_{cano}^{(0)} \left( {\cal D} 
= \frac{D} { \sqrt{ \sigma L} } \right)
\end{eqnarray}
where the scaling function
is determined by the integral transform
\begin{eqnarray}
\int_0^{+\infty} \frac{ d{\cal D} }{ {\cal D}^2 } 
e^{- \frac{ q^2 } { {\cal D}^2 } } {\cal P}_{cano}^{(0)} \left( {\cal D} \right)
&& =  \int_1^{+\infty} dv  
 \frac{ q  \cosh q  v \cosh q  (v-1)    - \sinh q  }
{ 2 \sinh^3 q  v} \nonumber \\
&& +   \int_{1/2}^1 dv
 \frac{  q  (2 v-1) \cosh q  v \cosh q  (1-v) 
-  \sinh q  (2 v-1)   }
{  2 \sinh^3 q v}
\end{eqnarray}
In particular, there is a logarithmic divergence 
in the limit $q \to 0$
\begin{eqnarray}
\int_0^{+\infty} \frac{ d{\cal D} }{ {\cal D}^2 } 
e^{- \frac{ q^2 } { {\cal D}^2 } } {\cal P}_{cano}^{(0)} \left( {\cal D} \right)
\opsimeq_{q \to 0} \frac{1}{2} \ln \frac{1}{q}
\end{eqnarray}
that yields the linear behavior near the origin
\begin{eqnarray}
 {\cal P}_{cano}^{(0)} \left( {\cal D} \right)
\oppropto_{{\cal D} \to 0} {\cal D}
\end{eqnarray}
in contrast with the essential singularity found
above for the microcanonical ensemble (\ref{microcald}).
As a consequence, the averaged
 correlation of the two surface magnetizations
decays algebraically in $L$  
\begin{eqnarray}
\left[ m^s_1 m^s_L \right]_{cano}^{scaling}
 \equiv \int_0^{+\infty} d {\cal D} P_{cano}( {\cal D} ) e^{ -
 {\cal D}  \frac{  \sqrt{ \sigma L} } {2}  }
 \oppropto_{L \to \infty} \frac{1}{L}
\label{averagedmsmscano} 
\end{eqnarray}
in contrast with the stretched exponential behavior
found for the microcanonical ensemble (\ref{averagedmsmsmicro}).
This big difference between the two ensembles
may be understood as follows :
in the microcanonical ensemble where $U(L)=0$,
the requirement to have a small $D$
is equivalent to the requirement to have a small
amplitude $A=U_{max}-U_{min}$, which is a very
strong constraint for a Brownian trajectory
that leads to an essential singularity for
the probability of small $D$. On the contrary, in the canonical ensemble,
the requirement to have a small $D=A-U(L)$
is not equivalent to the requirement of a small amplitude $A$ :
the freedom in $U(L)$ allows to have a large amplitude $A$ provided that
$U(L)$ is of the same order : the samples having a small $D$
are thus those having $U_{min} \sim 0$ ( probability $1/\sqrt{L}$)
and $U_{max} \sim U(L)$ ( probability $1/\sqrt{L}$)
and thus the measure of these samples is of order $1/L$,
in agreement with the above result (\ref{averagedmsmscano}).

Finally, we note that the decay as $1/L$ 
for the averaged correlation of the two surface magnetizations(\ref{averagedmsmscano}) in the canonical ensemble
is similar to the decay found
in \cite{rgfinitesize} for the averaged spin-spin end-to-end correlation
in the canonical ensemble. 
We will discuss the relation
between these two observables in more details in the next Section.

\section{ Gap and End-to-end spin-spin correlation in the two ensembles }

\label{gapcorre}

In this Section, we consider the end-to-end spin-spin correlation
and the gap in finite samples.
In contrast with the surface magnetizations
studied in previous Sections, these observables
cannot be written in closed form in terms of all the random couplings.
However, they have been studied in details in the 
canonical ensemble via the real-space renormalization approach
\cite{rgfinitesize}, whose predictions are in good agreement 
with the numerical data \cite{rgfinitesize,dharyoung}. 
The aim in this Section is to derive the corresponding results for
the microcanonical ensemble.

\subsection{ Gap and End-to-end spin-spin correlation in a given sample }

We refer the reader to the papers \cite{daniel,rgfinitesize}
for a detailed description of the real-space renormalization approach.
Here, we will only recall its results concerning
 the interpretation of the end-to-end spin-spin correlation
$C(L)$ (\ref{defcl}) and of the gap $\Delta(L)$ (\ref{defgapl})
for a given sample \cite{rgfinitesize} :
for a given realization of the random potential $U(x)$ for $0 \leq x \leq L$,
the gap $\Delta(L)$ is asymptotically determined by
the last renormalized ascending barrier $G$ via 
\begin{eqnarray}
G=-\ln \Delta (L) 
\end{eqnarray}
and the correlation $C(L)$ is asymptotically determined by
the variable $\Lambda=G-U(L)+U(0)$ via 
\begin{eqnarray}
\Lambda=-\ln C (L) 
\end{eqnarray}
In particular, the RSRG predicts a very simple relation
between the gap and the correlation in each given sample
\begin{eqnarray}
G-\Lambda= U(L)-U(0)
\end{eqnarray}
This prediction has been confirmed by the numerical studies
at the critical point,
in the canonical sample where $(G-\Lambda)/\sqrt{L}$ 
was found to converge towards a Gaussian distribution
\cite{rgfinitesize,dharyoung}, as well as in the 
microcanonical sample where $(G-\Lambda)/\sqrt{L}$ 
was found to converge towards a delta distribution
\cite{dharyoung}.

We refer the reader to \cite{daniel,rgfinitesize}
for a full theoretical justification of these results.
However, it is interesting to mention here
 that the RSRG prediction
for $C(L)$ { \it in the presence of a fixed boundary condition }
exactly corresponds to the saddle-point analysis
of Section \ref{saddlems}
for the surface magnetizations.
Indeed, if the spin $\sigma_L=1$ is fixed, it cannot be decimated,
and the last ascending barrier is constrained to contain this end-spin :
the variable $G$ is thus determined by the minimum via $G=U(L)-U_{min}$
and the variable $\Lambda$ reads $\Lambda= U_{min}-U(0)$,
in agreement with the saddle-point analysis
of Section \ref{saddlems} for the surface magnetization $m_1^s$.
When the boundary spins are both free, the gap and the correlation
are not in general simple functions of the end-values $U(0)$ and $U(L)$
and of the extrema $U_{min}$ and $U_{max}$ of the random potential,
but are determined by the properties of the last stage
of the renormalization procedure as we now recall.

\subsection{ Last stage of the renormalization procedure }

In the RSRG approach \cite{daniel}, one needs at a given scale $\Gamma$
the measure $B^{\pm}_{\Gamma} (F,l)$
of the ascending (resp. descending)  Brownian 
paths that goes from $U(0)$ to $U(l)=F$ (resp. $U(l)=-F$)
with $F \geq \Gamma$ and with no return of more than $\Gamma$,
i.e. for two arbitrary points $0<x_1<x_2<l$, the potential has to satisfy
$U(x_1)-U(x_2)<\Gamma$ (resp. $U(x_2)-U(x_1)<\Gamma$) \cite{us_landscape,
us_energysinai}.
In Laplace transform with respect to $l$ these measure read
\cite{daniel}
\begin{eqnarray}
 B^{\pm}_{\Gamma} (F;p) \equiv
\int_0^{+\infty} dl e^{-p l} B^{\pm}_{\Gamma} (F,l)
&& = \theta(F-\Gamma)  e^{ \pm \frac{\mu}{2} \Gamma } \frac{q}{\sinh q \Gamma}
e^{ -(F-\Gamma) (\mp \frac{\mu}{2} +q \coth q \Gamma )}
\end{eqnarray}
where $q=\sqrt{p+\mu^2/4}$ and $\sigma=1$ to simplify the notations in this Section. For the boundaries, one also needs
the measure $E^{\pm}_{\Gamma} (F,l)$
of the ascending (resp. descending)  Brownian 
paths that goes from $U(0)$ to $U(l)=F$ 
where $F \geq 0$ is the maximum 
(resp. $U(l)=-F$ is the minimum)
with no return of more than $\Gamma$.
In Laplace transform with respect to $l$ these measure read
\cite{rgfinitesize,us_landscape,
us_energysinai}
\begin{eqnarray}
 E^{\pm}_{\Gamma} (F;p) && = \theta(F) 
e^{ -F (\mp \frac{\mu}{2} +q \coth q \Gamma )}
\end{eqnarray}

The joint distribution $P_L(G,\Lambda,U)$ of the variables
$G=-\ln \Delta(L)$, $\Lambda=-\ln C(L)$
and of the end-point $U_L=U$ (with $U(0)=0$)
is then determined by the last stage of the RSRG  \cite{rgfinitesize}
as
\begin{eqnarray}
P_L(G,\Lambda,U) = \int dl_1 dl_2 dl_3 d\Lambda_R d \Lambda_L
 E^-_G(\Lambda_R) B^+_G (G) E^-_G (\Lambda_L) \delta(L-(l_1+l_2+l_3)) \delta (\Lambda-(\Lambda_L+\Lambda_R)) \delta(U- (G-\Lambda)  )
\end{eqnarray}
i.e. in Laplace with respect to $L$
\begin{eqnarray}
{\hat P}(G,\Lambda,U;p) = \theta(G) \theta(\Lambda)
 \Lambda
e^{ - \Lambda ( \frac{\mu}{2} +q \coth q G )} 
e^{  \frac{\mu}{2} G } \frac{q}{\sinh q G}
 \delta(U- (G-\Lambda)  )
\end{eqnarray}

\subsection{ Canonical ensemble }

In the canonical ensemble, the joint distribution
$P_L^{(\mu)}(G,\Lambda)$ has for Laplace transform
\begin{eqnarray}
\int_0^{+\infty} dL e^{-p L} 
\left[  P_L^{(\mu)}(G,\Lambda) \right]_{cano} = \int dU {\hat P}(G,\Lambda,U;p)
=  \theta(G) \theta(\Lambda)
 \frac{q \Lambda }{\sinh q G} 
e^{  \frac{\mu}{2} (G-\Lambda) } e^{ - \Lambda q \coth q G } 
\end{eqnarray}
which indeed coincides with the result (45) of \cite{rgfinitesize}.
We refer to \cite{rgfinitesize} for a detailed analysis
of this result, and only quote in the following some important
results in order to compare them with the microcanonical case below.

\subsubsection{ At criticality $\mu=0$ }

At criticality, the law for the gap alone has for Laplace transform
\begin{eqnarray}
\int_0^{+\infty} dL e^{-p L} 
\left[  P_L^{\mu=0}(G) \right]_{cano} 
=  \theta(G) 
 \frac{ \sinh \sqrt{p} G  }{\sqrt{p}   \cosh^2 \sqrt{p} G  }  
\end{eqnarray}
that corresponds after Laplace inversion to
the series given in Eq (57) in \cite{rgfinitesize}.
In particular, the probability distribution of
the rescaled variable $g=G/\sqrt{L}$
presents the essential singularity at the origin
\begin{eqnarray}
{\cal P}_{cano}^{\mu=0}(g) \oppropto_{g \to 0} \frac{1}{g^3} 
e^{- \frac{\pi^2}{4 g^2} }   
\end{eqnarray}
 that determines the decay with $L$ of
the averaged gap \cite{rgfinitesize}
\begin{eqnarray}
\left[ \Delta(L) \right]_{cano}^{scaling} \oppropto_{L \to \infty} 
L^{1/6} e^{- \frac{3}{2} \left( \frac{\pi^2}{2} L \right)^{1/3} }
\label{meangapcano}
\end{eqnarray}

On the other hand, the law for the correlation alone
has for Laplace transform at criticality
\begin{eqnarray}
\int_0^{+\infty} dL e^{-p L} 
\left[  P_L^{(\mu=0)}(\Lambda) \right]_{cano} 
=   \theta(\Lambda)
\sqrt{p} \Lambda  \int_0^{+\infty} dG
 \frac{ 1 }{\sinh \sqrt{p} G}  e^{ - \Lambda \sqrt{p} \coth \sqrt{p} G } 
=   \theta(\Lambda)
\Lambda K_0(\sqrt{p} \Lambda)
\end{eqnarray}
which corresponds after Laplace inversion to the simple result
for the rescaled variable $\lambda=\Lambda/{\sqrt L}$
\begin{eqnarray}
{\cal P}_{cano}^{\mu=0}(\lambda)   
=  \theta(\lambda)
\frac{\lambda}{2} e^{- \frac{\lambda^2}{4 } }
\end{eqnarray}
in excellent with numerical data \cite{rgfinitesize}.
As a consequence, 
the averaged correlation decays as a power of $L$ \cite{rgfinitesize}
\begin{eqnarray}
\left[  C(L) \right]_{cano}^{scaling}   
\oppropto_{L \to \infty} \frac{1}{L}
\end{eqnarray}

So in the canonical ensemble, at criticality,
the gap and the spin-spin correlation have very different behaviors.
In particular there is an essential singularity at the origin only for the gap
and not for the correlation, and thus the decays
for the averaged values are very different.

\subsubsection{ Ordered phase $\mu=-2 \delta>0$  }

In the ordered phase, the variable $\Lambda$ is simply the sum of two independent barriers
$(U(0)-U_{min})$ and $(U_{max}-U(L))$ against the bias \cite{rgfinitesize}
\begin{eqnarray}
P^{(\mu>0)}_{cano} (\Lambda) \simeq \mu^2 \Lambda e^{ -\mu \Lambda}
\label{plambdaorderedcano}
\end{eqnarray}
whereas the variable $G= U(L)+\Lambda$ is asymptotically distributed with
the Gaussian distribution of $U(L)$ alone \cite{rgfinitesize}
\begin{eqnarray}
P^{(\mu>0)}_{cano} (G) \simeq \frac{1}{ \sqrt{4 \pi L} } e^{- \frac{ (G-\mu L)^2}{4 L} } 
\label{pgorderedcano}
\end{eqnarray}

\subsubsection{ Disordered phase $\mu= -2 \delta<0$  }

In the disordered phase, the probability distribution of the variable $G$ 
reads \cite{rgfinitesize}
\begin{eqnarray}
P^{(\mu<0)}_{cano} (G) \simeq (L \mu^2) \vert \mu \vert e^{-\vert \mu \vert G }
e^{- (L \mu^2) e^{-\vert \mu \vert G } }
\label{distrigdisorderedcano}
\end{eqnarray}
that can be interpreted \cite{rgfinitesize} as the distribution of the maximal barrier
among $N \sim L \mu^2$ independent variables drawn with the exponential
distribution $\vert \mu \vert e^{-\vert \mu \vert G }$ of barriers against the drift.
whereas the variable $\Lambda= -U(L)-G$ is asymptotically distributed with
the Gaussian distribution of $U(L)$ alone \cite{rgfinitesize}
\begin{eqnarray}
P^{(\mu<0)}_{cano} (\Lambda) \simeq \frac{1}{ \sqrt{4 \pi L} }
 e^{- \frac{ (\Lambda- \vert \mu \vert L)^2}{4 L} } 
\label{distrilambdadisorderedcano}
\end{eqnarray}

\subsection{ Microcanonical ensemble }

\subsubsection{ At criticality $\mu=0$ }

At criticality $\mu=0$, 
we obtain that the 
the joint distribution
$P_L^{(\mu=0)}(G,\Lambda)$ is determined by the Laplace transform
\begin{eqnarray}
\int_0^{+\infty} dL e^{-p L} 
\left[  \frac{ P_L^{(\mu=0) }(G,\Lambda) }{ \sqrt{ 4 \pi 
 L }} \right]_{micro} 
= {\hat P}(G,\Lambda,U=0;p)
=   \delta(\Lambda- G  )
 \frac{\sqrt{p} G}{\sinh \sqrt{p} G} e^{ -  \sqrt{p} G \coth \sqrt{p} G } 
\end{eqnarray}
So here $G$ and $\Lambda$ are identical, and 
the rescaled variable
\begin{eqnarray}
g \equiv \frac{G}{  \sqrt{L} }  = \frac{\Lambda}{  \sqrt{L} }   
\end{eqnarray}
is distributed with the distribution ${\cal P}_{micro} (g)$
determined by the integral transform
\begin{eqnarray}
\frac{1}{\sqrt \pi} \int_0^{+\infty} \frac{dg}{g} {\cal P}_{micro} (g)
e^{- \frac{q}{g^2} }
=   \frac{ \sqrt q }{\sinh \sqrt q } e^{ -  \sqrt q  \coth \sqrt q  } 
\end{eqnarray}
In particular, the following divergence near $q \to -\pi^2$ 
\begin{eqnarray}
\frac{1}{\sqrt \pi} \int_0^{+\infty} \frac{dg}{g} {\cal P}_{micro} (g)
e^{- \frac{q}{g^2} }
\oppropto_{q \to \pi^2} \frac { 1}{q+ \pi^2}
  e^{  \frac{2 \pi^2 }{q+ \pi^2} }
\end{eqnarray}
yields the following essential singularity
 for the probability distribution
${\cal P}_{micro} (g)$ at small argument
\begin{eqnarray}
{\cal P}_{micro} (g)
\oppropto_{g \to 0} \frac{1}{g^{3/2} } 
e^{- \frac{\pi^2}{g^2} + \frac{2 \pi \sqrt 2}{g} }
\end{eqnarray}
This behavior at small $g$ of the scaling function
characterizes
 the rare events having a 
gap $\Delta_L=e^{-g \sqrt L } $ (or 
equivalently a correlation $C_L=e^{-g \sqrt L } $) of order one,
which are expected to dominate the mean values
of the gap and of the correlation.
The dependence in $L$ of these mean values may thus be estimated by a 
saddle point method which yields
\begin{eqnarray}
\left[ \Delta_L \right]_{micro}^{scaling}  
= \left[ C_L \right]_{micro}^{scaling}
&& \simeq \int_0^{+\infty} dg {\cal P}_{micro} (g)
e^{- g \sqrt{L} }  \\
&& \oppropto_{L \to \infty} L^{-1/6} e^{ - \frac{3}{2} ( 2 \pi^2 L)^{1/3}
+ 2 ( 2 \pi^2 L)^{1/6}   }
\end{eqnarray}
So here we obtain that there is a sub-leading
power $L^{1/6}$ { \it in the exponential }
with respect to the leading decay as $E^{- c L^{1/3} }$,
in contrast with the canonical ensemble (\ref{meangapcano}).

\subsubsection{ Ordered phase $\mu=-2 \delta>0$  }

In the ordered phase, the variable $\Lambda$ is again the sum of two independent barriers
$(U(0)-U_{min})$ and $(U_{max}-U(L))$ against the bias,
as in the canonical ensemble (\ref{plambdaorderedcano}).
This distribution of $\Lambda$ now completely determines
the probability distribution of the variable $G= \mu L+\Lambda$,
in contrast with the Gaussian distribution 
of the canonical ensemble (\ref{plambdaorderedcano}).
In particular, the inequality $G>\mu L$ indicates that the average
of the gap $\Delta(L)=e^{-G}$ will be governed by
the exponent ${\tilde \nu}=1$ instead of the exponent $\nu=2$,
for the reasons given around Equation (\ref{nutildefromnu})
concerning the case of the averaged surface magnetization.

\subsubsection{ Disordered phase $\mu=-2 \delta<0$  }

In the disordered phase, the distribution of the variable $G$ 
will be asymptotically the same as in the canonical ensemble
(\ref{distrigdisorderedcano}),
since the interpretation \cite{rgfinitesize} as the the maximal barrier
among $N $ independent barriers against the drift
still holds in the microcanonical ensemble.
However, the distribution of the variable $\Lambda$
is now completely determined by the relation $\Lambda= \vert \mu \vert L+G$,
in contrast with the Gaussian distribution (\ref{distrigdisorderedcano})
of the canonical case.
In particular, the inequality $\Lambda> \vert \mu \vert L$ indicates that the average
of the spin-spin correlation $C(L)=e^{-\Lambda}$ will be governed by
the exponent ${\tilde \nu}=1$ instead of the exponent $\nu=2$,
for the reasons given around Equation (\ref{nutildefromnu})
concerning the case of the averaged surface magnetization.

\subsection{ Comparison between $C(L)$ and $m^s_1 m^s_L$ 
at the critical point }

In the ordered phase, where the two end spins have a spontaneous magnetization,
it is clear that the limit $C(\infty)$ corresponds to the product 
of two independent
end-point magnetizations of an infinite system \cite{rgfinitesize}.
Moreover, it was found numerically \cite{dharyoung}
that the observables $C(L)$ and $( m^s_1 m^s_L)$ 
were actually still closely related 
at the critical point :
the limit distributions 
were found to be very close in the microcanonical ensemble,
but different at large argument in the canonical ensemble \cite{dharyoung}.
On the other hand, the difference at small argument
is much smaller in the canonical ensemble,
and the average values $[C(L)]_{cano}$
and $[m^s_1 m^s_L]_{cano}$ were found to be almost indistinguishable
numerically in the canonical ensemble \cite{dharyoung}.
It is thus interesting to discuss these questions here.

From the RSRG approach, it is clear that if 
the positions $x_{min}$ and $x_{max}$ of the minimum $U_{min}$
and $U_{max}$ satisfy the order $0\leq x_{min}<x_{max} \leq L$,
then the largest ascending barrier is simply given by
the amplitude $G=U_{max}-U_{min}$. Then the variable
$\Lambda$ for the spin-spin correlation
is given by 
$\Lambda=(U_{max}-U(L))+(U(0)-U_{min}) $
and thus exactly coincides with the variable $D=A(L)-U(L)$
(\ref{defdl}) determining the correlation of the two surface magnetizations.
In particular, in the ordered phase, 
all samples satisfy $0<x_{min}<x_{max}<L$ in the asymptotic limit
$L \to \infty$,
and the variables $\Lambda=D$ are the sum of
two independent barriers against the bias.
At the critical point, half of the samples 
satisfy the constraint $x_{min}<x_{max}$,
and thus we obtain that {\it for one half of the samples } at criticality,
both in the canonical and in the microcanonical ensemble
we have the identity $\Lambda=D$
that yields the identity 
of the two rescaled variables $(\ln C(L))/\sqrt{L}=(\ln m^s_1 m^s_L)/\sqrt{L}$.
On the other hand, for the other half of the samples which have
$x_{max}<x_{min}$, the largest ascending barrier 
is smaller than the amplitude $G<U_{max}-U_{min}$
and thus $\Lambda<D$ are two different variables.
In conclusion, the identity $\Lambda=D$ for half of the samples 
could explain the similarities found numerically
in \cite{dharyoung} for their probability distributions.

Concerning the mean values, the interpretation
given after Equation (\ref{averagedmsmscano})
for the decay as $1/L$ of the averaged correlation
of the surface magnetizations in the canonical ensemble 
yields that the relevant rare events precisely satisfy the order
$0\leq x_{min} < x_{max} \leq L$. As a consequence, 
we expect that the leading term in $1/L$ of
the averaged spin-spin correlation exactly coincides
with the leading term in $1/L$ of the 
averaged correlation
of the surface magnetizations in the canonical ensemble,
in agreement with the numerical finding
that both quantities were almost indistinguishable \cite{dharyoung}.

\section{Conclusions}

\label{secconclusion}

The example of the RTFIC considered in this paper
shows that the question of the ensemble dependence for the
critical properties of disordered systems is rich and instructive.
Indeed, even if the two ensembles are expected to be equivalent in the
thermodynamic limit, their finite-size properties may nevertheless
be quite different. 

At criticality, even if the appropriate rescaled variables
for all observables considered here are the same in both ensembles,
the probability distributions may be quite different
in the two ensembles. In particular, 
some asymptotic behavior may be given by two different
power-laws in the two ensembles
(as in the case of the surface magnetization), 
or even by a power law in one ensemble
and by an essential singularity in the other ensemble
(as in the case of the end-to-end correlations). 
As a consequence, the decay with $L$ of averaged observables
can be drastically different in the two ensembles : for instance,
the end-to-end correlations decay as a power law in the canonical ensemble,
but as a stretched exponential in the microcanonical ensemble.

Off criticality, the probability distributions of rescaled variables
 involve the exponent $\nu=2$ in both ensembles, but averaged observables
are again governed by rare samples, whose measures
can be very sensitive to the microcanonical constraint.
As a consequence, the critical properties of a given averaged observable may 
be governed by two different exponents in the two ensembles. 
For instance, we have shown via an exact path-integral method that the
averaged surface magnetization in the disordered phase
was governed by the exponent $\nu=2$ in the canonical ensemble
and by the exponent ${\tilde \nu}=1$ in the microcanonical ensemble,
in agreement with the interpretation of \cite{igloi98}, 
and in contrast with the interpretation of \cite{dharyoung}.
We have moreover concluded from the domains of definition of
probability distributions of other observables in the scaling regime,
that similarly, (a) in the disordered phase, the averaged correlation of the two surface magnetizations and the end-to-end spin-spin correlation are governed by the exponent $\nu=2$ in the canonical ensemble and by the exponent $\tilde \nu=1$ in the microcanonical ensemble (b) in the ordered phase, the averaged gap is governed by the exponent $\nu=2$ in the canonical ensemble and by the exponent $\tilde \nu=1$ in the microcanonical ensemble.

As a final remark, let us mention the very recent preprint \cite{refaeldaniel},
where the RSRG method is generalized to study the end-to-end energy-energy correlation $C_E(L)$
(in the canonical ensemble), and where averaged observables are again governed by rare events.

 \begin{acknowledgments}

It is a pleasure to thank T. Garel and J. Houdayer for useful discussions.

\end{acknowledgments}

\appendix

\section{Joint distribution of the amplitude $A$ and the end-point $U$ }

\label{appjointau}

The probability $G_{[U_a,U_b]}(U,U_0;l)$
for a random walk with bias $F_0=\mu \sigma$ 
to go from $U_0=0 $ to $U$
during $L$ 
in the presence of two absorbing boundaries at $U=U_a<0$ and
$U=U_b>0$ has for Laplace transform 
\cite{us_landscape,us_energysinai}  
\begin{eqnarray}
\hat{G}_{[U_a,U_b]}^{\mu}(U,p|U_0) = e^{ \frac{\mu}{2} (U-U_0) } \frac{
\sinh q (U-U_a) \sinh q (U_b- U_0)}
{ \sigma q \sinh q (U_b-U_a)} 
 \qquad \text{if} ~~ U_a \leq U \leq U_0 \\
\hat{G}_{[U_a,U_b]}^{\mu}(U,p|U_0) =  e^{ \frac{\mu}{2} (U-U_0) } 
\frac{\sinh q (U_b- U) \sinh q (U_0-U_a)}
{ \sigma q \sinh q (U_b-U_a)} 
 \qquad \text{if} ~~ U_0 \leq U \leq U_b 
\label{laplacef}
\end{eqnarray}
where
\begin{eqnarray}
q \equiv \sqrt{\frac{p}{\sigma}+\frac{\mu^2}{4} }
\end{eqnarray}

The joint probability of the end-point $U$,
of the minimum value $U_a$ and of the maximal value $U_b$
when starting at $U_0=0$
has thus for Laplace transform
\begin{eqnarray}
&& {\hat P}^{(\mu)}(U,U_a,U_b;p)  = 
- \partial_{ U_a} \partial_{ U_b}
{ \hat G}^{(F_0)}_{[U_a,U_b]}(U,p \vert 0) \\
&& = e^{ \frac{\mu}{2} U }
\frac{q}{ \sigma }  \frac{
 \sinh qU \sinh q (U_b+U_a)
- 2 \cosh q U \sinh q U_a \sinh q U_b-   }
{  \sinh^3 q (U_b-U_a)}  
\label{laplacepuaub}
\end{eqnarray}

Finally, the joint probability of the amplitude $A_L=U_{max}-U_{min}=U_b-U_a$
and the end-point $U$ when starting at $U_0=0$
has for Laplace transform
\begin{eqnarray}
 {\hat P}^{(\mu)}(A,U;q)  && = 
 \int_{-\infty}^{+\infty} dU_a \int_{-\infty}^{+\infty} dU_b 
\theta(U_a<0<U_b)
\theta(U_a<U<U_b) {\hat P}^{(\mu)}(U,U_a,U_b) \delta( A-(U_b-U_a) ) \\
&& = \theta(A>\vert U \vert) \int_{max(-A,U-A)}^{min(0,U)} dU_a  
{\hat P}^{(\mu)}(U,U_a,U_a+A)
\end{eqnarray}
This finally leads to the result (\ref{loijointeulamplilaplace}) given in the text.

\section{Useful properties of Bessel functions }

The series expansion of Bessel function at small argument reads
\begin{eqnarray}
K_{iq}(z) && = \frac{\pi}{2 i \sinh \pi q}
\sum_{k=0}^{+\infty} \frac{\left( \frac{z}{2} \right)^{2k}}{k! } 
 \left[\frac{\left( \frac{z}{2} \right)^{-iq}}{\Gamma(-iq+k+1)}
-\frac{\left( \frac{z}{2} \right)^{iq}}{\Gamma(iq+k+1)} \right]
\label{kiqzto0}
 \end{eqnarray}
whereas the asymptotic behavior at large argument is given by
\begin{eqnarray}
K_{\nu}(z) \opsimeq_{z \to \infty}
\sqrt{ \frac{\pi}{2 z }  } e^{-z} 
\left[  \sum_{k=0}^{n} \frac{1}{ k! (2 z)^k}
\frac{\Gamma(\nu+k+\frac{1}{2} )}{\Gamma(\nu-k+\frac{1}{2} )}
+.. \right]
\label{knulargez}
 \end{eqnarray}

A useful integral is
\begin{eqnarray}
\int_0^{+\infty} dp p^{\frac{\mu}{2}-1} e^{-p} K_{iq} \left( 2 \sqrt{
\frac{p}{\sigma} }  \right)
= \frac{\sqrt \sigma }{2} e^{\frac{1 }{ 2 \sigma }}
\Gamma \left( \frac{ \mu+1+iq}{2} \right)
\Gamma \left( \frac{ \mu+1-iq}{2} \right)
W_{-\frac{\mu}{2} , i \frac{q}{2} } 
\left(  \frac{1 }{  \sigma } \right)
\label{whit}
 \end{eqnarray}
where $W$ is the Whittaker function.
For the special case $\mu=0$, it simplifies into
\begin{eqnarray}
W_{0 , r } 
\left( z \right) = \sqrt{ \frac{z}{\pi} } K_{r} \left( \frac{z}{2} \right)
\label{whitzero}
 \end{eqnarray}

\end{document}